\documentclass{emulateapj}
\usepackage{graphicx}
\usepackage{epstopdf}
\usepackage{color}

\newcommand{\Mjup}{\mbox{$M_\mathrm{Jup}$}}
\newcommand{\Msun}{\mbox{$M_{\odot}$}}

\begin{document}
\title{Imaging Extrasolar Giant Planets}
\author{Brendan P. Bowler\altaffilmark{1,2} 
\\ }
\email{bpbowler@astro.as.utexas.edu}

\altaffiltext{1}{McDonald Observatory and the University of Texas at Austin, Department of Astronomy, 2515 Speedway, Stop C1400, Austin, TX 78712}
\altaffiltext{2}{McDonald Prize Fellow.}

\begin{abstract}
High-contrast adaptive optics imaging is a powerful technique to probe the architectures of planetary systems from the outside-in and survey the atmospheres of self-luminous giant planets.  Direct imaging has rapidly matured over the past decade and especially the last few years with the advent of high-order adaptive optics systems, dedicated planet-finding instruments with specialized coronagraphs, and innovative observing and post-processing strategies to suppress speckle noise.  This review summarizes recent progress in high-contrast imaging with particular emphasis on observational results, discoveries near and below the deuterium-burning limit, and a practical overview of large-scale surveys and dedicated instruments.  I conclude with a statistical meta-analysis of deep imaging surveys in the literature.  Based on observations of 384 unique and single young ($\approx$5--300~Myr) stars spanning stellar masses between 0.1--3.0~\Msun, the overall occurrence rate of 5--13~\Mjup \ companions at orbital distances of 30--300~AU is 0.6$^{+0.7}_{-0.5}$\% assuming hot-start evolutionary models.  The most massive giant planets regularly accessible to direct imaging are about as rare as hot Jupiters are around Sun-like stars.  Dividing this sample into individual stellar mass bins does not reveal any statistically-significant trend in planet frequency with host mass: giant planets are found around 2.8$^{+3.7}_{-2.3}$\% of BA stars, $<$4.1\% of FGK stars, and $<$3.9\% of M dwarfs.  Looking forward, extreme adaptive optics systems and the next generation of ground- and space-based telescopes with smaller inner working angles and deeper detection limits will increase the pace of discovery to ultimately map the demographics, composition, evolution, and origin of planets spanning a broad range of masses and ages.

\end{abstract}

\section{Introduction}{\label{sec:intro}}

\begin{figure*}
  \vskip -1.2 in
  \hskip -.3 in
  \resizebox{7.7in}{!}{\includegraphics{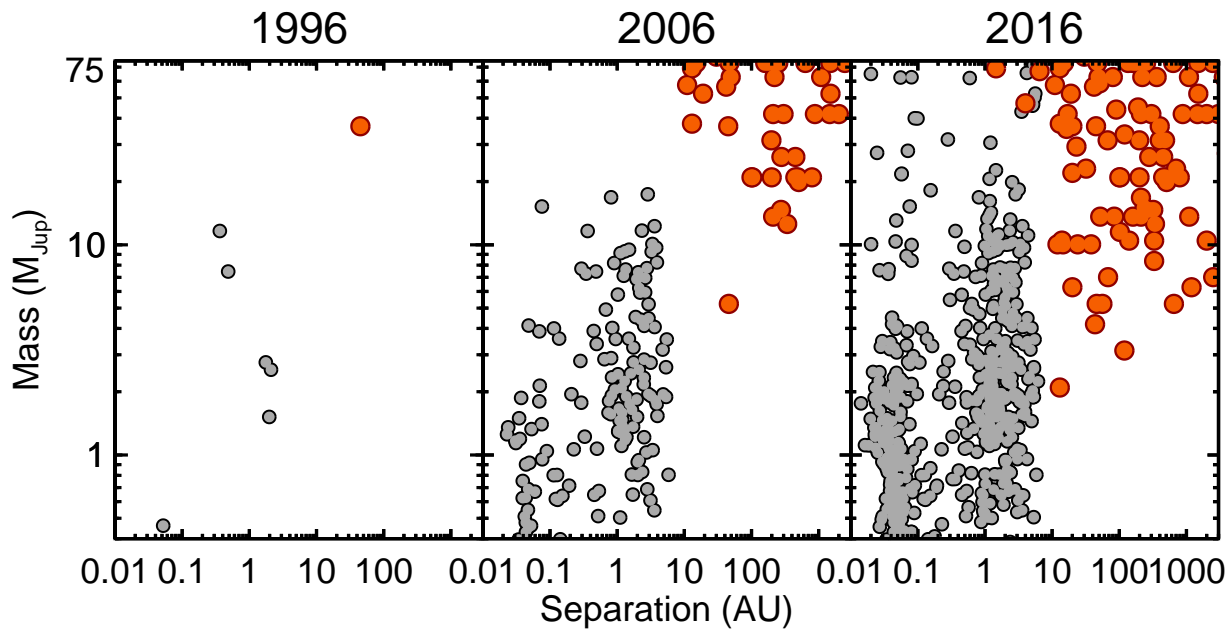}}
  \vskip -1.1 in
  \caption{Substellar companions discovered via radial velocities (gray circles) and direct imaging (red circles) as of 1996, 2006, and 2016.  Over this twenty year period the number of directly imaged companions below 10~\Mjup \ has steadily increased from one (2M1207--3932~b in 2004) to over a dozen.  The surprising discovery of planetary companions at extremely wide separations of hundreds to thousands of AU has expanded the architecture of planetary systems to over five orders of magnitude.  Note that the radial velocity planets are minimum masses ($m$sin$i$) and the directly imaged companion masses are inferred from evolutionary models.   RV-detected planets are from exoplanets.org (\citealt{Wright:2011cqa}; \citealt{Han:2014hn}) and are supplemented with a compilation of RV-detected brown dwarfs from the literature.  Imaged companions are from \citet{Deacon:2014ey} together with other discoveries from the literature.  \label{fig:mass_sma} } 
\end{figure*}

Over the past two decades the orbital architecture of giant planets has expanded from a single order of magnitude in the Solar System (5--30 AU) to over five orders of magnitude among extrasolar planetary systems (0.01--5000 AU; Figure~\ref{fig:mass_sma}).  
High-contrast adaptive optics (AO) imaging has played a critical role in this advancement by probing separations beyond $\sim$10~AU and masses $\gtrsim$1~\Mjup.  Uncovering planetary-mass objects at 
hundreds and thousands of AU has fueled novel theories of planet formation and migration, inspiring a more complex framework for the origin of giant planets in which multiple mechanisms (core accretion, dynamical scattering, disk instability, and cloud fragmentation) operate on different timescales and orbital separations.  In addition to probing unexplored orbital distances, imaging entails directly capturing photons that originated in planetary atmospheres, providing unparalleled information about the initial conditions, chemical composition, internal structure, atmospheric dynamics, photospheric condensates, and physical properties of extrasolar planets.  
These three science goals --- the architecture, formation, and atmospheres of gas giants --- represent the main motivations to directly image and spectroscopically characterize extrasolar giant planets.

This pedagogical review summarizes the field of direct imaging in the era leading up to and transitioning towards extreme adaptive optics systems, the \emph{James Webb Space Telescope}, \emph{WFIRST},  and the thirty meter-class telescopes.  This ``classical'' period of high-contrast imaging spanning approximately 2000 to 2015 has set the stage and baseline expectations for the next generation of instruments and telescopes that will deliver ultra-high contrasts and reach unprecedented sensitivities.  
In addition to the first images of \emph{bona fide} extrasolar planets, this early phase experienced a number of surprising discoveries including planetary-mass companions orbiting brown dwarfs; planets on ultra-wide orbits beyond 100~AU; enigmatic (and still poorly understood) objects like the optically-bright companion to Fomalhaut; and unexpectedly red, methane-free, and dust-rich atmospheres at low surface gravities.  Among the most important results has been the gradual realization that massive planets are exceedingly rare on wide orbits; only a handful of discoveries have been made despite thousands of hours spent on hundreds of targets spanning over a dozen surveys.  Although dismaying, these null detections provide valuable information about the efficiency of planet formation and the resulting demographics at wide separations.  Making use of mostly general, non-optimized facility instruments and early adaptive optics systems has also led to creative observing strategies and post-processing solutions for PSF subtraction.

Distinguishing giant planets from low-mass brown dwarfs is a well-trodden intellectual exercise (e.g., \citealt{Oppenheimer:2000vo}; \citealt{Basri:2006kv}; \citealt{Chabrier:2007ty}; \citealt{Chabrier:2014up}).  
Except in the few rare cases where the architectures or abundance patterns of individual systems offer clues about a specific formation route, untangling the origin of imaged planetary-mass companions must necessarily be addressed as a population and in a statistical manner.  This review is limited in scope to self-luminous companions detected in thermal emission at near- and mid-infrared wavelengths (1--5~$\mu$m) with masses between $\approx$1--13~\Mjup \ with the understanding that multiple formation routes can probably produce objects in this ``planetary'' mass regime (see Section~\ref{sec:dbl}).
Indeed, the separations regularly probed in high-contrast imaging surveys--- typically tens to hundreds of AU--- 
lie beyond the regions in protoplanetary disks containing the highest surface densities of solids where core accretion operates most efficiently (e.g., \citealt{Andrews:2015vzb}).
Direct imaging has therefore predominantly surveyed the wide orbital distances where alternative formation and migration channels like disk instability, cloud fragmentation, and planet-planet scattering are most likely to apply.
In the future, the most efficient strategy to detect even smaller super-Earths and terrestrial worlds close to their host stars will be in reflected light from a dedicated space-based optical telescope.

By focusing on the optimal targets, early discoveries, largest surveys, and statistical results, this observationally-oriented overview aims to complement recent reviews on giant planet formation (\citealt{Chabrier:2007ty}; \citealt{Helled:2013et}; \citealt{Chabrier:2014up}; \citealt{Helling:2014hb}), atmospheric models (\citealt{Marley:2007uc}; \citealt{Helling:2008gs}, \citealt{Allard:2012fp}; \citealt{Marley:2015bj}), evolutionary models, (\citealt{Burrows:2001wq}; \citealt{Fortney:2009jt}), observational results (\citealt{Absil:2009dja}; \citealt{Lagrange:2014gp}; \citealt{Bailey:2014dc}; \citealt{Helling:2014fx}; \citealt{Madhusudhan:2014wu}; \citealt{Quanz:2015fg}; \citealt{Crossfield:2015jd}), and high contrast imaging instruments and speckle suppression techniques (\citealt{Guyon:2006jp}; \citealt{Beuzit:2007tj}; \citealt{Oppenheimer:2009gh}; \citealt{Biller:2008kk}; \citealt{Marois:2010hs}; \citealt{Traub:2010vo}; \citealt{Mawet:2012il}; \citealt{Davies:2012ds}).

\begin{figure*}
  \vskip -1.2 in
  \hskip -.2 in
  \resizebox{7.3in}{!}{\includegraphics{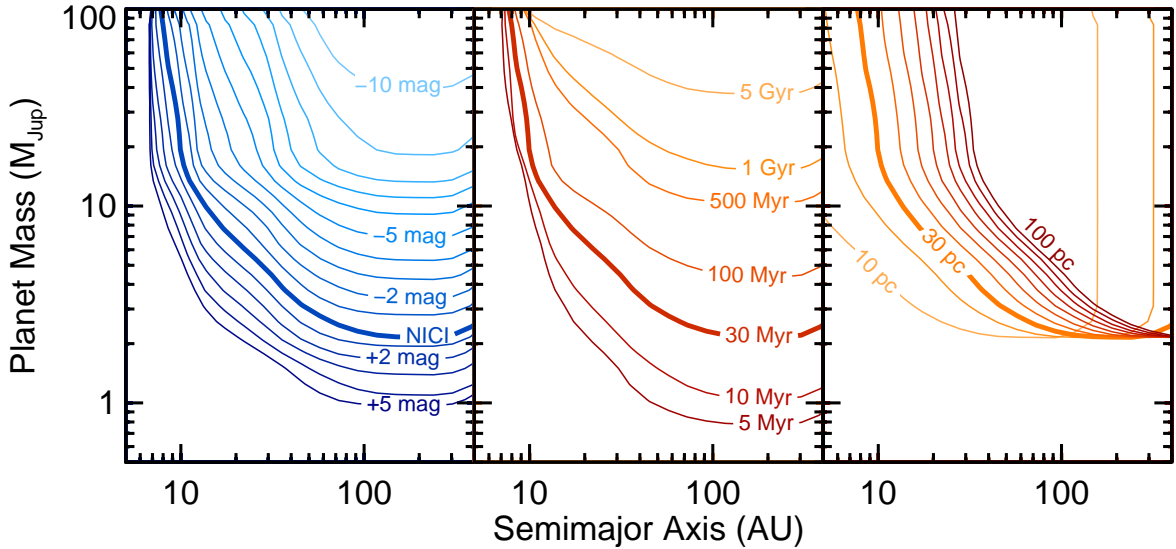}}
  \vskip -1 in
  \caption{The influence of contrast (left), age (middle), and distance (right) on mass sensitivity to planets.  
  The bold curve in each panel shows the 50\% sensitivity contour based on the median 
  NICI contrast from \citet{Biller:2013fu} for a 30~Myr K1 star at 30~pc.  The left panel shows the effect of 
  increasing or decreasing the fiducial contrast curve between --10 magnitudes to +5 magnitudes.  Similarly, the middle and
  right panels show changes to the fiducial age spanning 5~Myr to 5~Gyr and distances spanning 10~pc to 100~pc.
  Planet absolute magnitudes depend steeply on mass and age. As a result, a small gain in contrast in the 
  brown dwarf regime corresponds to a large
  gain in limiting mass, but the same contrast gain in the planetary regime translates into a much smaller gain in mass.  
  Mass sensitivity is particularly sensitive to stellar age, while closer distances mean smaller physical separations can be studied.
   \label{fig:contrast_sensitivitylimits} } 
\end{figure*}

\section{Optimal Targets for High-Contrast Imaging}

Planets radiatively cool over time by endlessly releasing the latent heat generated during their formation and gravitational contraction.  
Fundamental scaling relations for the evolution of brown dwarfs and giant planets
can be derived analytically with basic assumptions of a polytropic equation of state and
degenerate electron gas (\citealt{Stevenson:1991ie}; \citealt{Burrows:1993kt}).  
Neglecting the influence of lithium burning, deuterium burning, and atmospheres, 
which act as partly opaque wavelength-dependent boundary conditions,
substellar objects with different masses cool in a similar monotonic fashion over time:

\begin{equation}
L_\mathrm{bol} \propto t^{-5/4} M^{5/2}.
\end{equation}

\noindent Here $L_\mathrm{bol}$ is the bolometric luminosity, $t$ is the object's age, and $M$
is its mass.  

This steep mass-luminosity relationship means that luminosity tracks are compressed in the 
brown dwarf regime ($\approx$13--75~\Mjup) and  
fan out in the planetary regime with significant consequences for
high-contrast imaging.  
A small gain in contrast in the brown dwarf regime results in a large gain in the 
limiting detectable mass, whereas the same contrast gain in the planetary regime
has a much smaller influence on limiting mass (Figure~\ref{fig:contrast_sensitivitylimits}).  
It is much more difficult, for example, to improve sensitivity from
10~\Mjup \ to 1~\Mjup \ than from 80~\Mjup \ to 10~\Mjup.
Moreover, sensitivity to low masses and close separations is highly dependent on 
a star's youth and proximity.  
In terms of limiting detectable planet mass, observing younger and closer stars is 
equivalent to improving speckle suppression or integrating for longer.
Note that in a contrast-limited regime 
the absolute magnitude of the host star is also important.  
The same contrast around low-mass stars and brown dwarfs corresponds to lower limiting masses compared to
higher-mass stars.  

Young stars are therefore attractive targets for two principal reasons: planets are their most luminous at early ages, and the relative contrast between young giant planets and their host stars is lower than at older ages because stellar luminosities plateau on the main sequence while planets and brown dwarfs continue to cool, creating a luminosity bifurcation.  For example, evolutionary models predict the $H$-band contrast between a 5 \Mjup \ planet orbiting a 1 \Msun \ star to be $\approx$25~mag at 5 Gyr but only $\approx$10~mag at 10 Myr (\citealt{Baraffe:2003bj}; \citealt{Baraffe:2015fw}).  At old ages beyond $\sim$1~Gyr, 1--10 \Mjup \ planets are expected to have effective temperatures between 100--500~K and cool to the late-T and Y spectral classes with near-infrared absolute magnitudes $\gtrsim$18~mag (\citealt{Dupuy:2013ks}).

Below are overviews of the most common classes of targets in direct imaging surveys highlighting the
scientific context, strengths and drawbacks, and observational results for each category.

\subsection{Young Moving Group Members}

In principle, younger stars make better targets for imaging planets.  In practice, the youngest T Tauri stars reside in star-forming regions beyond 100~pc.  At these distances, the typical angular scales over which high-contrast imaging can probe planetary masses translate to wide physical separations beyond $\sim$20--50 AU (with some notable exceptions with non-redundant aperture masking and extreme AO systems).  Moreover, these extremely young ages of $\sim$1--10~Myr correspond to timescales when giant planets may still be assembling through core accretion and therefore might have lower luminosities than at slightly later epochs (e.g., \citealt{Marley:2007bf}; \citealt{Molliere:2012go}; \citealt{Marleau:2013bh}).  On the other hand, the closest stars to the Sun probe the smallest physical scales but their old ages of $\sim$1--10 Gyr mean that high contrast imaging only reaches brown dwarf masses.

\begin{figure*}
  \vskip -1.2 in
  \hskip -.6 in
  \resizebox{7.8in}{!}{\includegraphics{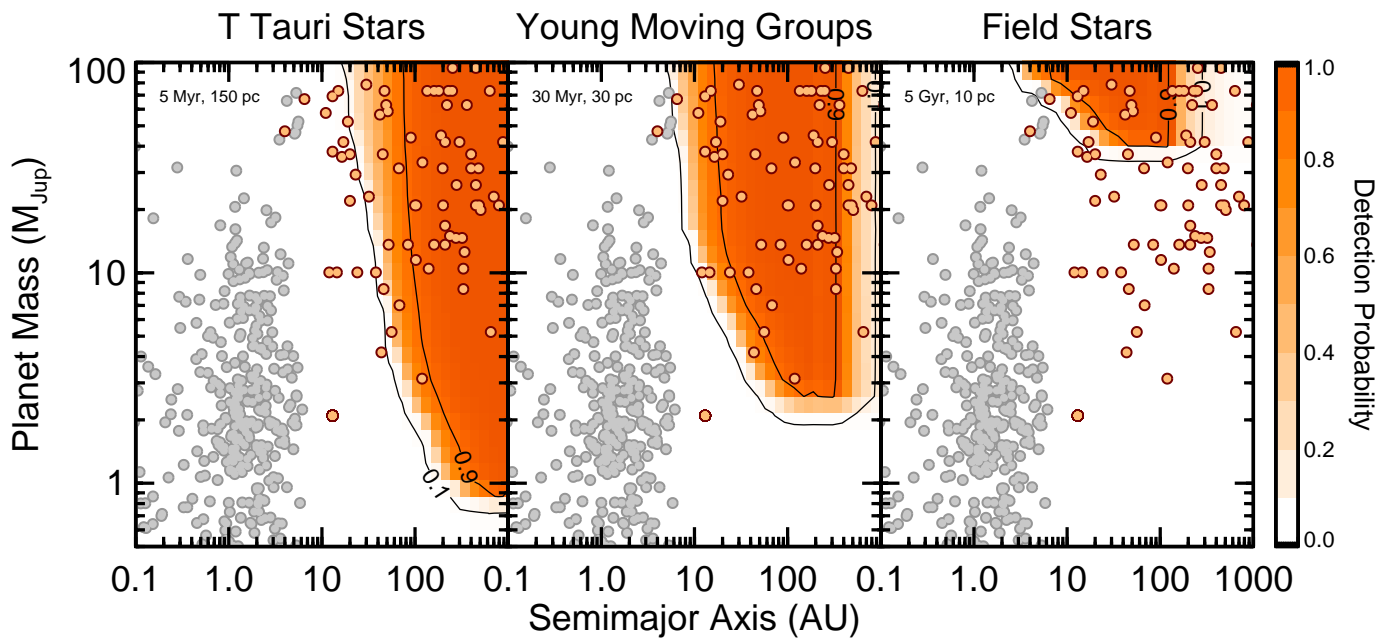}}
  \vskip -1.2 in
  \caption{Typical sensitivity maps for high-contrast imaging observations of T Tauri stars (5~Myr at 150~pc),
  young moving group members (30~Myr at 30~pc), and field stars (5~Gyr at 10~pc).  Young moving group members are ``Golidlocks targets''---
  not too old, not too distant.  Black curves denote 10\% and 90\% contour
  levels assuming circular orbits, Cond hot-start evolutionary models (\citealt{Baraffe:2003bj}), and the median NICI contrast curve
  from \citet{Biller:2013fu}.  Gray and orange circles are RV- and directly imaged companions, respectively (see Figure~\ref{fig:mass_sma}). \label{fig:ymg_sensitivitylimits} } 
\end{figure*}

Young moving groups--- coeval, kinematically comoving associations young stars and brown dwarfs in the solar neighborhood--- represent a compromise in age  ($\approx$10--150~Myr) and distance ($\approx$10--100 pc) between the nearest star forming regions and field stars (Figure~\ref{fig:ymg_sensitivitylimits}; see \citealt{Zuckerman:2004ku}, \citealt{Torres:2008vq}, and \citealt{Mamajek:2016ik}).  One distinct advantage they hold is that their members span a wide range of masses and can be used to age date each cluster from lithium depletion boundaries and isochrone fitting (e.g., \citealt{Bell:2015gw}; \citealt{Herczeg:2015bp}).  As a result, the ages of these groups are generally much better constrained than those of isolated young stars.  For these reasons young moving group members have emerged as the primary targets for high-contrast imaging planet searches over the past decade (e.g., \citealt{Chauvin:2010hm}; \citealt{Biller:2013fu}; \citealt{Brandt:2014hc})

Identifying these nearby unbound associations of young stars is a difficult task.  Each moving group's $UVW$ space velocities cluster closely together with small velocity dispersions of $\approx$1--2~km s$^{-1}$ but individual stars in the same group can be separated by tens of parsecs in space and tens of degrees across the sky.  $UVW$ kinematics can be precisely determined if the proper motion, radial velocity, and parallax to a star are known.  Incomplete knowledge of one or more of these parameters (usually the radial velocity and/or distance) means the $UVW$ kinematics are only partially constrained, making it challenging to unambiguously associate stars with known groups.  Historically, most groups themselves and new members of these groups were found with the aid of the Tycho Catalog and \emph{Hipparcos}, which provided complete space velocities for bright stars together with ancillary information pointing to youth such as infrared excess from IRAS; X-ray emission from the $Einstein$ or $ROSAT$ space observatories; strong H$\alpha$ emission; and/or \ion{Li}{1} $\lambda$6708 absorption.  As a result, most of the faint low-mass stars and brown dwarfs have been neglected.

\begin{figure}
  \vskip -.45 in
  \hskip -1.1 in
  \resizebox{6.3in}{!}{\includegraphics{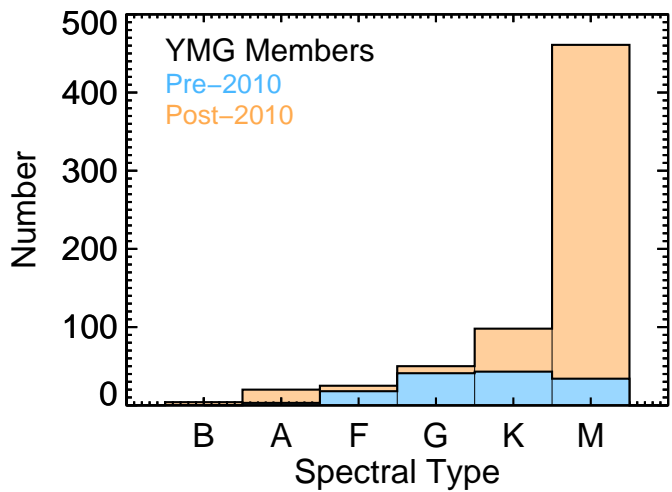}}
  \vskip -1.4 in
  \caption{The census of members and candidates of young moving groups.  Prior to 2010 the M dwarf members were
  largely missing owing to their faintness and lack of parallax measurements from $Hipparcos$.  
  Concerted efforts to find low-mass members over the past few years have filled in this population and 
  generated a wealth of targets for dedicated direct imaging planet searches.  \label{fig:newmgmembers} } 
\end{figure}

In recent years the population of ``missing'' low-mass stars and brown dwarfs in young moving groups has been increasingly uncovered as a result of large all-sky dedicated searches (Figure \ref{fig:newmgmembers}; \citealt{Shkolnik:2009dx}; \citealt{Lepine:2009ey}; \citealt{Schlieder:2010gk}; \citealt{Kiss:2010cb}; \citealt{Rodriguez:2011gb}; \citealt{Schlieder:2012gj}; \citealt{Schlieder:2012gu}; \citealt{Shkolnik:2012cs}; \citealt{Malo:2013gn}; \citealt{Moor:2013cy}; \citealt{Rodriguez:2013fv}; \citealt{Malo:2014dk}; \citealt{Gagne:2014gp}; \citealt{Riedel:2014ce}; \citealt{Kraus:2014ur}; \citealt{Gagne:2015ij}; \citealt{Gagne:2015dc}; \citealt{Binks:2015bu}).  Parallaxes and radial velocities are generally not available for these otherwise anonymous objects, but by adopting the $UVW$ kinematics of known groups, it is possible to invert the problem and predict a distance, radial velocity, and membership probability.  Radial velocities are observationally cheaper to acquire en masse compared to parallaxes, so membership confirmation has typically been accomplished with high-resolution spectroscopy.  The exceptions are for spectroscopic binaries, which require multiple epochs to measure a systemic velocity, and rapidly rotating stars with high projected rotational velocities ($v$sin$i$), which produce large uncertainties in radial velocity measurements.  The abundance of low-mass stars in the field means that some old interlopers will inevitably share similar space velocities with young moving groups.  These must be distilled from bona fide membership lists on a case-by-case basis (\citealt{Barenfeld:2013bf}; \citealt{Wollert:2014go}; \citealt{Janson:2014gz}; \citealt{Mccarthy:2014jp}; \citealt{Bowler:2015ch}).
 
The current census of directly imaged planets and companions near the deuterium-burning limit are listed in Table~\ref{tab:planets}.  Many of these host stars are members of young moving groups.  $\beta$~Pic, 51~Eri, and possibly TYC~9486-927-1 are members of the $\beta$~Pic moving group (\citealt{Zuckerman:2001go}; \citealt{Feigelson:2006tz}; \citealt{Deacon:2016dg}).  HR~8799 and possibly $\kappa$~And are thought to be members of Columba (\citealt{Zuckerman:2011bo}).  2M1207--3932 is in the TW~Hydrae Association (\citealt{Gizis:2002je}).  GU~Psc and 2M0122--2439 are likely members of the AB Dor moving group (\citealt{Malo:2013gn}; \citealt{Naud:2014jx}; \citealt{Bowler:2013ek}). 
AB~Pic, 2M0103--5515, and 2M0219--3925 are in Tuc-Hor (\citealt{Song:2003hh}; \citealt{Delorme:2013bo}; \citealt{Gagne:2015ij}), though the masses of their companions are somewhat uncertain and may not reside in the planetary regime.  In addition, the space motion of VHS~1256--1257 is well-aligned with the $\beta$~Pic or possibly AB~Dor moving groups (\citealt{Gauza:2015fw}; \citealt{Stone:2016fz}), but the lack of lithium in the host indicates the system is older and may be a kinematic interloper.

The number of moving groups in the solar neighborhood is still under debate, but at least five are generally considered to be well-established: the TW Hydrae Association, $\beta$~Pic, Tuc-Hor, Carina, and AB Dor.  Other associations have been proposed and may constitute real groups which formed together and are useful for age-dating purposes, but may require more scrutiny to better understand their size, structure, physical nature, and relationship to other groups.  \citet{Mamajek:2016ik} provide a concise up-to-date census of their status and certitude.  Soon, micro-arcsecond astrometry and parallaxes from $Gaia$ will dramatically change the landscape of nearby young moving groups by readily identifying overlooked groups, missing members, and even massive planets on moderate orbits.

\subsection{T Tauri Stars, Herbig Ae/Be Stars, and \\ Transition Disks}

Despite their greater distances ($\approx$120--150~pc), the extreme youth ($\approx$1--10~Myr) of 
T Tauri stars and their massive counterparts, Herbig Ae/Be stars, 
in nearby star-forming regions like Taurus, the Sco-Cen complex, and $\rho$ Oph
have made them attractive targets to search 
for planets with direct imaging and probe the earliest stages of planet formation when gas giants are still assembling 
(\citealt{Itoh:2008ta}; \citealt{Ireland:2011id}; \citealt{Mawet:2012fz}; \citealt{Janson:2013ke}; 
\citealt{LaFreniere:2014dj}; \citealt{Daemgen:2015fp}; \citealt{Quanz:2015fg}; \citealt{Hinkley:2015dga}).  

One of the most 
surprising results from these efforts has been the unexpected discovery of planetary-mass companions
on ultra-wide orbits at several hundred AU from their host stars
(Table~\ref{tab:planets}).   
These wide companions pose challenges to canonical theories of planet formation
via core accretion and disk instability and may instead represent the tail end of brown
dwarf companion formation, perhaps as opacity-limited fragments of 
turbulent, collapsing molecular cloud cores (e.g., \citealt{Low:1976wt}; \citealt{Silk:1977il}; \citealt{Boss:2001vw}; \citealt{Bate:2009br}).

Many (and perhaps most) of these young planetary-mass companions harbor accreting circum-planetary disks, which
provide valuable information about mass accretion rates, circum-planetary disk structure, 
formation route, and the moon-forming capabilities of young planets.
Accretion luminosity is partially radiated in line emission, making H$\alpha$ a potentially
valuable tracer to find and characterize protoplanets (\citealt{Sallum:2015ej}).  
For example,  \citet{Zhou:2014ct} find that up to 50\% of 
the accretion luminosity in the $\approx$15~\Mjup \ companion GSC~6214-210~B 
is emitted at H$\alpha$.  
Searching for these nascent protoplanets has become a leading motivation to achieve AO correction
in the optical and is actively being carried out with MagAO (\citealt{Close:2014kt}).
Deep sub-mm observations with ALMA 
have opened the possibility of measuring the masses of these subdisks
(\citealt{Bowler:2015hx}) and possibly even indirect identification via gas kinematics (\citealt{Perez:2015jn}).  
Larger disks may be able to be 
spatially resolved and a dynamical mass for the planet may be measured from Keplerian motion.  

The relationship between protoplanetary disks and young planets is also being explored 
in detail at these extremely young ages.  In particular, transition disks--- 
young stars whose spectral energy distributions indicate they host disks with 
large optically thin cavities generally depleted of dust (e.g., see reviews by 
\citealt{Williams:2011js}, \citealt{Espaillat:2014hh}, \citealt{Andrews:2015vzb}, and \citealt{Owen:2016id})--- 
have been used as signposts to search for embedded protoplanets.   
This approach has been quite fruitful, resulting in the discovery of companions within these gaps
spanning the stellar (CoKu Tau 4: \citealt{Ireland:2008kj}; 
HD~142527: \citealt{Biller:2012cb}, \citealt{Close:2014kt}, \citealt{Rodigas:2014bm}, \citealt{Lacour:2015uv}),
brown dwarf (HD~169142: \citealt{Biller:2014ft}, \citealt{Reggiani:2014dj}), 
and planetary (LkCa~15: \citealt{Kraus:2012gk}, \citealt{Ireland:2014jm}, \citealt{Sallum:2015ej}; 
HD~100546: \citealt{Quanz:2013ii}, \citealt{Currie:2014hn}, \citealt{Quanz:2015dz}; \citealt{Currie:2015jk}, \citealt{Garufi:2016vt})
mass regimes using a variety of techniques.

However, environmental factors can severely complicate the interpretation of these detections.
Extinction and reddening, accretion onto and from circum-planetary disks, extended emission from accretion streams, 
and circumstellar disk sub-structures seen in scattered light can result in false alarms, degenerate interpretations,
and large uncertainties in the mass estimates of actual companions.  
T Cha offers a cautionary example; \citet{Huelamo:2011hx} discovered a candidate substellar companion
a mere 62~mas from the transition-disk host star with aperture masking interferometry, 
but additional observations did not show orbital motion as expected for a real companion.  Additional 
modeling indicates
that the signal may instead be a result of scattering by grains in the outer disk or possibly even noise in the data
(\citealt{Olofsson:2013cx}; \citealt{Sallum:2015gm}; \citealt{Cheetham:2015hg}).
This highlights an additional complication with aperture masking: because model fits  
to closure phases usually consist of binary models with two or more point sources, it can difficult to discern 
actual planets from other false positives.  In these situations the astrometric detection of orbital motion is essential to confirm young 
protoplanets embedded in disks.

Other notable examples of ambiguous candidate protoplanets at wider separations include 
FW~Tau~b, an accreting low-mass companion to the Taurus binary FW~Tau~AB orbiting at a projected
separation of 330~AU (\citealt{White:2001ic}; \citealt{Kraus:2014tl}), 
and TMR-1C, a faint, heavily extincted protoplanet candidate located $\approx$1400~AU 
from the Taurus protostellar binary host TMR-1AB
showing large-amplitude photometric variability and circumstantial evidence of a dynamical ejection 
(\citealt{Terebey:1998co}; \citealt{Terebey:2000jq}; \citealt{Riaz:2011fj}; \citealt{Riaz:2013fg}).
Follow-up observations of both companions suggest they may instead be low-mass stars or brown dwarfs
with edge-on disks (\citealt{PetrGotzens:2010hj}; \citealt{Bowler:2014dk}; \citealt{Kraus:2015fx}; \citealt{Caceres:2015hg}), 
underscoring a few of the difficulties that arise when interpreting candidate protoplanets at
extremely young ages.

Altogether the statistics of planets orbiting the youngest T Tauri stars from direct imaging are still 
fairly poorly constrained.  Quantifying this occurrence rate is important because it can be compared with the
same values at older ages to determine the \emph{evolution} of this population.  Planet-planet
scattering, for example, implies an increase in the frequency of planets on ultra-wide orbits over time.
\citet{Ireland:2011id} found the frequency of 6--20~\Mjup \ companions from 
$\approx$200--500~AU to be $\sim$4$^{+5}_{-1}$\% in Upper Scorpius.  
Combining these results with those from \citet{Kraus:2008bh} and their own shallow imaging survey, 
\citet{LaFreniere:2014dj} find that the frequency of 5--40~\Mjup \ companions between
50--250~AU is $<$1.8\% and between 250--1000~AU is 4.0$^{+3.0}_{-1.2}$\%
assuming hot-start evolutionary models.
In future surveys it will be just as important to report nondetections together with new discoveries so this frequency
can be measured with greater precision.

\subsection{Brown Dwarfs}

Young brown dwarfs ($\approx$13--75~\Mjup) 
have low circum-substellar disk masses (\citealt{Mohanty:2013kl}; \citealt{Andrews:2013ku}) and
are not expected to host giant planets as frequently as stars.
Nevertheless, 
their low luminosities make them especially advantageous for high-contrast imaging 
because lower masses can be probed with contrast-limited observations. 

Several deep imaging surveys with ground-based AO or $HST$ have included brown dwarfs in their samples 
(\citealt{Kraus:2005tc}; \citealt{Ahmic:2007ju}; \citealt{Stumpf:2010es}; \citealt{Biller:2011hq}; 
\citealt{Todorov:2014fqa}; \citealt{Garcia:2015dra}).  
A handful of companions in the 5--15~\Mjup \ range have been discovered with direct imaging:
2M1207--3932~b (\citealt{Chauvin:2004cy}; \citealt{Chauvin:2005gg}),
2M0441+2301~b (\citealt{Todorov:2010cn}),
and possibly both FU~Tau~B (\citealt{Luhman:2009cx})
and VHS~1256--1257 (\citealt{Gauza:2015fw}) depending on their ages.
A few other low-mass companions 
(and in some cases the primaries themselves) to late-T and early-Y field brown dwarfs 
may also reside in the planetary regime depending on the system ages:
CFBDSIR~J1458+1013~B (\citealt{Liu:2011hb}), WISE~J1217+1626~B (\citealt{Liu:2012cy}), and 
WISE~J0146+4234~B (\citealt{Dupuy:2015dza}).

The low mass ratios of these systems ($q$$\approx$0.2--0.5) bear a closer resemblance to binary stars 
than canonical planetary systems ($q$$\lesssim$0.001), and the formation route of these very low-mass binaries 
is probably quite different than around stars (\citealt{Lodato:2005ef}).
High-order multiple systems with low total masses like 2M0441+2301 AabBab and VHS~1256--1257 ABb suggest
that cloud fragmentation can form objects in the planetary-mass domain (\citealt{Chauvin:2005gg}; 
\citealt{Todorov:2010cn}; \citealt{Bowler:2015en}; \citealt{Stone:2016fz}).
Continued astrometric monitoring of ultracool binaries will eventually yield orbital elements and 
dynamical masses for these intriguing systems to test
formation mechanisms (\citealt{Dupuy:2011ip}) and giant planet evolutionary models.

\subsection{Binary Stars}

Close stellar binaries ($\approx$0.1--5$''$) are generally avoided in direct imaging surveys. 
Multiple similarly-bright stars can confuse wavefront sensors, which are 
optimized for single point sources, and
deep coronagraphic imaging generally saturates nearby stellar companions.  
Physically, binaries carve out large dynamically-unstable regions that are inhospitable to planets 
and there is strong evidence that they inhibit planet formation 
by rapidly clearing or truncating protoplanetary disks (e.g., \citealt{Cieza:2009hb}; \citealt{Duchene:2010gb}; \citealt{Kraus:2012dh}).
Nevertheless, many planets have been found in binary systems in both S-type orbital configurations 
(a planet orbiting a single star; e.g., \citealt{Ngo:2015hn})
and P-type orbits (circumbinary planets; see \citealt{Winn:2015jt} for a recent summary).
Binaries are common products of star formation, 
so understanding how stellar multiplicity influences 
the initial conditions (protoplanetary disk mass and structure),
secular evolution (Kozai-Lidov interactions) 
and end products (dynamically relaxed planetary systems)
of planet formation has important consequences for the galactic census of exoplanets
(e.g., \citealt{Wang:2014jf}; \citealt{Kraus:2016wn}).

Several planetary-mass companions have been imaged around binary stars on wide circumbinary orbits:
ROXs~42B~b (\citealt{Kraus:2014tl}; \citealt{Currie:2014gp}), Ross~458~c (\citealt{Goldman:2010ct}; \citealt{Scholz:2010cy}), 
SR~12~C (\citealt{Kuzuhara:2011ic}), 
HD~106906~b (\citealt{Bailey:2014et}; \citealt{Lagrange:2016bh}), and VHS~1256--1257 (\citealt{Gauza:2015fw}; \citealt{Stone:2016fz}).
2M0103--5515~b (\citealt{Delorme:2013bo}) and FW Tau~b (\citealt{Kraus:2014tl}) orbit close, near-equal mass stellar binaries, but
the masses of the wide tertiaries are highly uncertain (\citealt{Bowler:2014dk}).

On the other hand, few imaged planets orbiting single stars also have wide stellar companions.  51~Eri Ab is orbited by the pair of
M dwarf binaries GJ~3305 AB (\citealt{Feigelson:2006tz}; \citealt{Kasper:2007dm}; \citealt{Montet:2015ky}) at $\sim$2000~AU.
Fomalhaut has two extremely distant stellar companions at $\approx$57 kAU and $\approx$158 kAU 
(\citealt{Mamajek:2013gzb}).
2M0441+2301 Bb orbits a low-mass brown dwarf and is part of a hierarchical quadruple system with
a distant star-brown dwarf pair at a projected separation of 1800 AU (\citealt{Todorov:2010cn}).

There has been little work comparing the occurrence rate of imaged planets in binaries and single stars.  
However, several surveys and post-processing techniques are now expressly focusing on binary systems and should clarify the statistical properties of planets
in these dynamically complicated arrangements (\citealt{Thalmann:2014bq}; \citealt{Rodigas:2015hq}; \citealt{Thomas:2015iga}).

\subsection{Debris Disks: Signposts of Planet Formation?}

Debris disks are extrasolar analogs of the asteroid and Kuiper belts in the Solar System.
They are the continually-replenished outcomes of cascading planetesimal collisions that result
in large quantities of transient dust heated by the host star.  Observationally, debris disks 
are identified from unresolved infrared or sub-mm excesses over stellar photospheric emission.
Deep observations spanning the optical, IR, and sub-mm can spatially resolved the largest and most luminous disks 
in scattered light and thermal emission to investigate disk morphology and grain properties.
Topical reviews by \citet{Zuckerman:2001ez}, \citet{MoroMartin:2008vi}, \citet{Wyatt:2008ht}, 
\citet{Krivov:2010er}, and \citet{Matthews:2014id} highlight 
recent theoretical and observational progress on the formation, modeling, and evolution of debris disks.

Debris disks are intimately linked to planets, which can stir planetesimals,
sculpt disk features, produce offsets between disks and their host stars, 
and carve gaps to form belts with spectral energy distributions showing multiple temperature components.
The presence of debris disks, and especially those with features indicative of a massive perturber, 
may therefore act as signposts for planets.  
The four directly imaged planetary systems $\beta$~Pic, HR~8799, 51~Eri, and HD~95086 
all possess debris disks, the latter three having multiple belts interior and exterior to the imaged planet(s).
This remarkably consistent configuration is analogous to the Solar System's architecture in 
which gas giants are flanked by (very low-level) zodiacal emission.

Anecdotal signs point to a possible correlation between disks and imaged planets
but this relationship has not yet been statistically validated.
There have been hints of a correlation between debris disks and
low-mass planets detected via radial velocities (\citealt{Wyatt:2012kn}; \citealt{Marshall:2014bp}),
but these were not confirmed in a recent analysis by \citet{MoroMartin:2015dk}.
Indeed, many stars hosting multi-component debris disks have now been targeted
with high-contrast imaging and do not appear to harbor massive planets 
(\citealt{Rameau:2013it}; \citealt{Wahhaj:2013iq}; \citealt{Janson:2013cjb}; \citealt{Meshkat:2015dh}). 
Given the high incidence of debris disks around main-sequence stars ($\gtrsim$16--20\%: 
\citealt{Trilling:2008ey}; \citealt{Eiroa:2013bk}),
with even higher rates at younger ages (\citealt{Rieke:2005hv}; \citealt{Meyer:2008vh}), 
any correlation of imaged giant planets and debris disks will be difficult to discern because 
the overall occurrence rate of massive planets on wide orbits is extremely low ($\lesssim$1\%; see Section~\ref{sec:occurrencerate}).
Perhaps more intriguing would be a subset of this sample with additional contextual clues, 
for example the probability of an imaged planet given a two-component debris disk compared to a diskless
control sample.

The Fabulous Four--- Vega, $\beta$~Pic, Fomalhaut, and $\epsilon$ Eridani--- host the brightest 
debris disks discovered by IRAS (\citealt{Aumann:1984bg}; \citealt{Aumann:1985zz}) 
and have probably been targeted more than any other stars with 
high-contrast imaging over the past 15 years, except perhaps for HR~8799.  Despite having similarly large and
luminous disks, their planetary systems are quite different and demonstrate
a wide diversity of evolutionary outcomes.

Fomalhaut's disk possesses a sharply truncated, offset, and eccentric ring about 140~AU in radius 
suggesting sculpting from a planet (\citealt{Dent:2000ix}; \citealt{Boley:2012fh}; \citealt{Kalas:2005ca}).  
A comoving optical source (``Fomalhaut b'') was discovered 
interior to the ring by \citet{Kalas:2008un} and appears to be orbiting on a highly inclined and eccentric orbit not
coincident with the ring structure (\citealt{Kalas:2013hpa}).  
The nature of this intriguing companion remains puzzling; it may be a low-mass planet with a large circum-planetary disk, a swarm of
colliding irregular satellites,  or perhaps
a recent collision of protoplanets (e.g., \citealt{Kalas:2008cs}; \citealt{Kennedy:2011ca}; \citealt{Kenyon:2014kf}).  
Massive planets have been ruled out from deep imaging down to 
about 20 AU (\citealt{Kalas:2008un}; \citealt{Kenworthy:2009hc}; \citealt{Marengo:2009de}; \citealt{Absil:2011jm}; 
\citealt{Janson:2012bn}; \citealt{Nielsen:2013jy}; \citealt{Kenworthy:2013bt}; 
\citealt{Currie:2012ef}; \citealt{Currie:2013iw}; \citealt{Janson:2015bh}).

Vega's nearly face-on debris disk is similar to Fomalhaut's in terms of its two-component structure 
comprised of warm and cold dust belts and wide gaps with orbital ratios $\gtrsim$10, possibly indicating the presence
of multiple low-mass planets (e.g., \citealt{Wilner:2002wl}; \citealt{Su:2005wx}; \citealt{Su:2013da}).
Deep imaging of Vega over the past 15 years has thus far failed to identify planets with detection 
limits down to a few Jupiter masses 
 (\citealt{Metchev:2003fv}; \citealt{Macintosh:2003in}; \citealt{Itoh:2006vp}; \citealt{Marois:2006df}; \citealt{Hinz:2006tx}; 
\citealt{Hinkley:2007hf}; \citealt{Heinze:2008fg}; \citealt{Janson:2011hu}; \citealt{Mennesson:2011ia}; \citealt{Janson:2015bh}).

$\beta$~Pic hosts an extraordinarily large, nearly edge-on disk spanning almost 2000 AU in radius (\citealt{Smith:1984wy}; \citealt{Larwood:2001wd}).   
Its proximity, brightness, and spatial extent make it one of the best-studied debris disks, showing signs of multiple belts (e.g., \citealt{Wahhaj:2003eh}),
asymmetries (e.g., \citealt{Kalas:1995jo}), molecular gas clumps (\citealt{Dent:2014br}), and 
an inner warp (\citealt{Heap:2000jr}) predicted to be caused by a close-in inclined planet (\citealt{Mouillet:1997ib}).
\citet{Lagrange:2009hq} uncovered a possible massive planet at $\sim$9~AU and, despite immediate follow-up 
(\citealt{Fitzgerald:2009fs}; \citealt{Lagrange:2009gg}), it was not until it reemerged on
the other side of the star that $\beta$~Pic~b was unambiguously confirmed (\citealt{Lagrange:2010fsa}).

$\epsilon$ Eridani is another particularly fascinating example of a nearby, relatively young K2 star hosting a bright
debris disk with spatially resolved ring structure and a warm inner component 
(\citealt{Greaves:1998kg}; \citealt{Greaves:2005vn}; \citealt{Backman:2009hk}).  
At 3.2~pc this star harbors the closest debris disk to the Sun, has a Jovian-mass planet 
detected by radial velocity and astrometric variations (\citealt{Hatzes:2000gr}; \citealt{Benedict:2006dr}), 
and possesses a long-term RV trend pointing to an additional long-period giant planet.  
Because of its favorable age and proximity, 
$\epsilon$ Eridani has been exhaustively imaged with adaptive optics on 
the largest ground-based telescopes in an effort to recover the known planets and search for others
(\citealt{Luhman:2002ed}; \citealt{Macintosh:2003in}; \citealt{Itoh:2006vp}; \citealt{Marengo:2006gj}; \citealt{Janson:2007cm}; 
\citealt{Lafreniere:2007cv}; \citealt{Biller:2007ht}; \citealt{Janson:2008gp}; \citealt{Heinze:2008fg}; 
\citealt{Marengo:2009de};  \citealt{Heinze:2010dm}; \citealt{Wahhaj:2013iq}; \citealt{Janson:2015bh}).
Together with long-baseline radial velocity monitoring, these deep observations have ruled out planets $>$3~\Mjup \ anywhere in this system.

\subsection{Field Stars and Radial Velocity Trends}

At the old ages of field stars ($\sim$1--10~Gyr), giant planets have cooled to 
late spectral types and
low luminosities where high-contrast imaging does not regularly reach 
the planetary-mass regime.
Nevertheless, several  surveys have focused on this population
because their proximity means very close separations can be probed and their old ages provide
information about potential dynamical evolution of substellar companions and giant planets over time
(\citealt{McCarthy:2004hl}; \citealt{Carson:2005ia}; \citealt{Carson:2006ez}; 
\citealt{Heinze:2010ko}; \citealt{Heinze:2010dm}; \citealt{Tanner:2010bp}; 
\citealt{Leconte:2010ed}).  

Of particular interest are stars showing low-amplitude, long-baseline radial velocity changes (Doppler ``trends''). 
These accelerations are regularly revealed in planet searches  and point to the existence of unseen stars, brown dwarfs, or giant planets on wide orbits.  High-contrast imaging is a useful
tool to diagnose the nature of these companions and, in the case of non-detections, rule out massive objects at 
wide projected separations (\citealt{Kasper:2007dm}; \citealt{Geissler:2007dg}; 
\citealt{Luhman:2002ed}; \citealt{Chauvin:2006hm}; \citealt{Janson:2009id}; \citealt{Jenkins:2010cz}; 
\citealt{Kenworthy:2009hc}; \citealt{Rodigas:2011gp}).  

When a companion is detected, its minimum mass can be inferred from the host star's acceleration ($\dot{v}$),
the distance to the system ($d$), and the angular separation of the companion ($\rho$) following \citet{Torres:1999gc} and
\citet{Liu:2002fx}:

\begin{equation}
M_\mathrm{comp} > 1.388 \times 10^{-5}  \left( \frac{d}{\mathrm{pc}} \ \frac{\rho}{''}   \right) ^2 \Big|  \frac{\dot{v}}{\mathrm{m} \ \mathrm{s^{-1}} \ \mathrm{yr^{-1}}} \Big|  \ M_{\odot}.
\end{equation}

\noindent The coefficient  is 0.0145 when expressed in \Mjup.
This equation assumes an instantaneous radial velocity slope, but longer baseline coverage 
or a change in the acceleration (``jerk'') can provide better constraints on a companion's mass and 
period (\citealt{Rodigas:2016cf}).  If a significant fraction of the orbit is measured
with both astrometry and radial velocities, simultaneous modeling of both data sets can yield a  
robust dynamical mass measurement.  
Perhaps the best example of this is from \citet{Crepp:2012eg}, who measured
the mass and three-dimensional orbit of the brown dwarf companion HR~7672~B,  
initially discovered by \citet{Liu:2002fx} based on an acceleration from the host star.

\begin{figure}
  \vskip -.5 in
  \hskip -0.7 in
  \resizebox{5.1in}{!}{\includegraphics{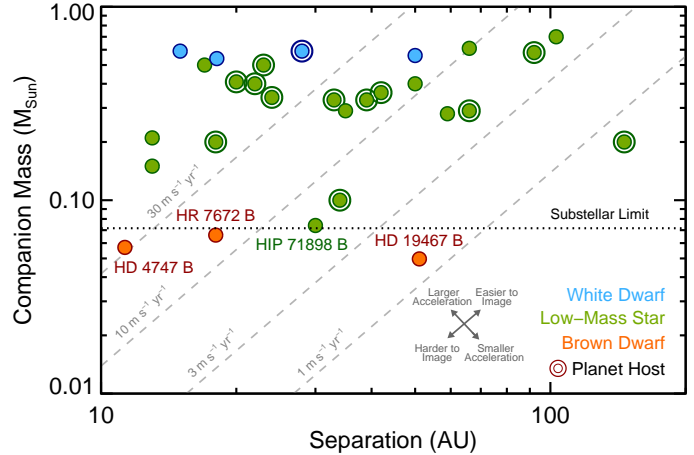}}
  \vskip -.7 in
  \caption{Imaged companions inducing shallow radial velocity trends on their host stars.  
  Blue, orange, and red circles are white dwarf companions, low-mass stellar companions, and substellar companions, respectively.   
  Concentric circles indicate the host star has a planetary system.
  Only three brown dwarf companions inducing shallow trends have been found: 
  HR~7672~B (\citealt{Liu:2002fx}), HD~19467~B (\citealt{Crepp:2014ce}), and HD~4747~B (\citealt{Crepp:2016ta}).
  Gray dashed lines show constant accelerations assuming circular orbits;
  the maximum host star acceleration is proportional to the companion mass and
inversely proportional to the square of the projected physical separation, so 
a 1~\Mjup \ planet at 10~AU 
will produce the same maximum acceleration as a 100~\Mjup \ low-mass star at 100 AU (namely
0.7~m s$^{-1}$ yr$^{-1}$).  See Table~\ref{tab:trends} for details on these systems.  \label{fig:trends} } 
\end{figure}

Many stellar and white dwarf companions have been discovered in this fashion but only
a few substellar companions have been found (Table~\ref{tab:trends}). 
Figure~\ref{fig:trends} shows the population of known companions inducing shallow RV trends on their host stars
and which have also been recovered with high resolution (and often high-contrast) imaging.
Most of these are M dwarfs with masses between $\sim$0.1--0.5~\Msun \ at separations of $\sim$10--100~AU.
This is primarily due to the two competing methods at play: at these old ages, 
direct imaging is insensitive to low 
masses and close separations, while small accelerations induced from wide-separation
and low-mass companions are difficult to measure even for long-baseline, precision radial velocity
planet searches.  
The TRENDS program (e.g., \citealt{Crepp:2012eg}; \citealt{Crepp:2014ce}) is the largest survey 
to combine these two methods and demonstrates the importance of both 
detections and non-detections to infer the population of planets on moderate orbits out to $\sim$20~AU (\citealt{Montet:2014fa}).

Dynamical masses of planets may eventually be measured by combing radial velocity monitoring of the host star and direct imaging,
effectively treating the system like a spatially resolved single-lined spectroscopic binary.  
Stellar jitter is a limiting factor at very young ages and at older ages the low luminosities of planets generally preclude
imaging.  The intermediate ages of moving group members may provide an adequate solution,
and at least one ambitious survey by \citet{Lagrange:2013gh} 
is currently underway to search for planets and long-term radial velocity trends
for this population.  Another solution is to image planets in reflected light at optical wavelengths, 
which requires a space-based telescope and 
coronagraph like \emph{WFIRST} (\citealt{Traub:2014ft}; \citealt{Spergel:2015wr}; 
\citealt{Brown:2015bj}; \citealt{Greco:2015ko}; \citealt{Robinson:2016gb}).
Similarly, astrometric accelerations can be used to identify and measure the masses of substellar companions when combined with 
high-contrast imaging.  This will be particularly relevant in the near-future with $Gaia$ as thousands of
planets are expected to be found from the orbital reflex motion of their host stars (\citealt{Sozzetti:2013de}; \citealt{Perryman:2014jra}).

Young stars in the field  not necessarily associated with coherent moving groups have also been 
popular targets for high-contrast imaging planet searches.  The advantage of this population is that they are numerous
and often reside at closer distances than actual members of young moving groups, but their ages and metallicities are generally highly uncertain so substellar companions uncovered with deep imaging can 
have a wide range of possible masses (e.g., \citealt{Mawet:2015kk}).
GJ~504 and $\kappa$~And are recent examples of young field stars with faint companions 
that were initially thought to have planetary masses 
(\citealt{Kuzuhara:2013jz}; \citealt{Carson:2013fw}) but which 
follow-up studies showed are probably older (\citealt{Hinkley:2013ko}; \citealt{Bonnefoy:2014dx}; \citealt{Fuhrmann:2015dk}; \citealt{Jones:2016hg}),
implying the companions are likely more massive and
reside in the brown dwarf regime.
Sirius is another example of a young massive field star extensively 
targeted with high-contrast imaging (\citealt{Kuchner:2000kf}; \citealt{BonnetBidaud:2008hk}; 
\citealt{Skemer:2011is}; \citealt{Thalmann:2011jx}; \citealt{Vigan:2015fsa}).
This system is particularly noteworthy for possible periodic astrometric perturbations to the orbit of its white dwarf companion Sirius B that may be caused
by an still-hidden giant planet or brown dwarf (\citealt{Benest:1995to}).

\section{The Masses of Imaged Planets}{\label{sec:masses}}

The masses of directly imaged planets are generally highly uncertain, heavily model-dependent,
and difficult to independently measure.  
Yet mass is fundamentally important to test models of giant planet formation and empirically calibrate substellar evolutionary models.
This Section describes how observables like bolometric luminosity, color, and absolute magnitude 
coupled with evolutionary models and semi-empirical quantities like age are used to infer the masses of planets.
Although no imaged planet has yet had its mass directly measured, there are several promising
routes to achieve this which will eventually enable rigorous tests of giant planet cooling models.

\begin{figure*}
  \vskip -1.2 in
  \hskip -.5 in
  \resizebox{8.4in}{!}{\includegraphics{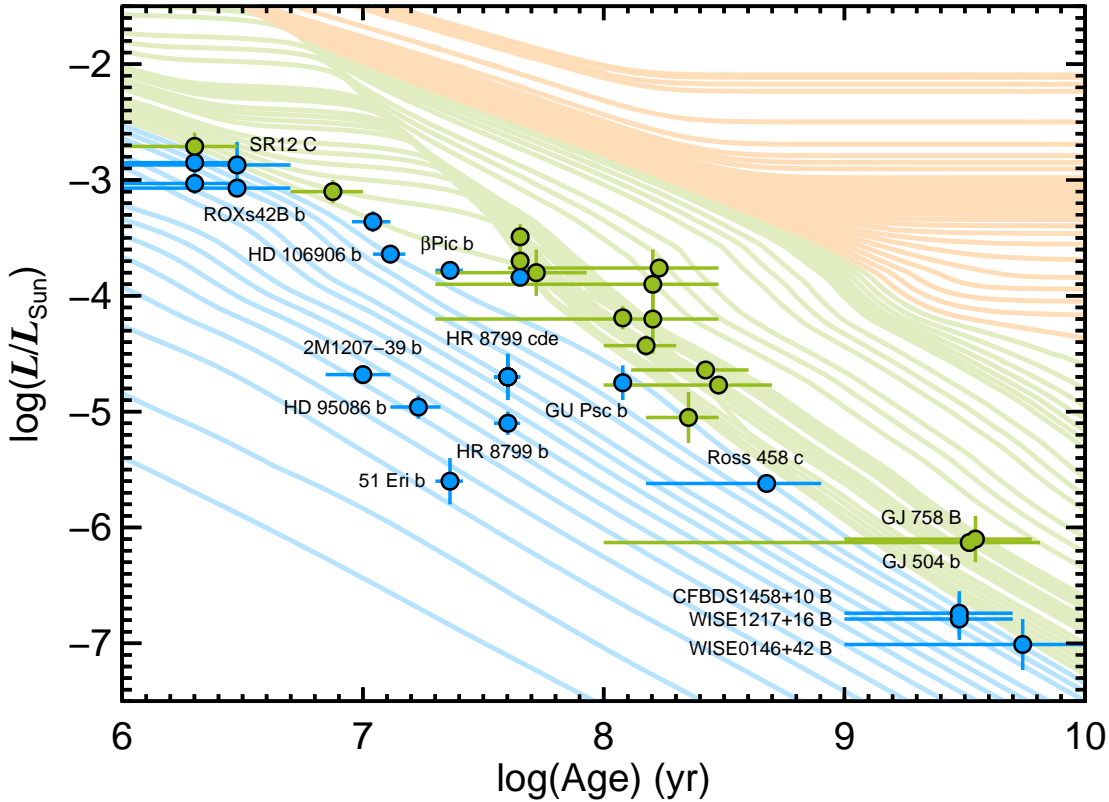}}
  \vskip -0.7 in
  \caption{The current census of companions in the brown dwarf (green) and planetary (blue) mass regimes that have both age and bolometric luminosity measurements from the compilation in Table~\ref{tab:planets}.  Many companions lie near the deuterium-burning limit while only a handful of objects are unambiguously in the planetary-mass regime.  Hot-start evolutionary models are from \citet{Burrows:1997jq}; orange, green, and blue tracks denote masses $>$80~\Mjup, 14--80~\Mjup, and $<$14~\Mjup. \label{fig:lum_age} } 
\end{figure*}

\subsection{Inferring Masses}

Like white dwarfs and brown dwarfs, giant planets
 cool over time so evolutionary models along with two physical parameters---
luminosity, age, effective temperature, or radius---
are needed to infer a planet's mass.  
Among these, luminosity and age are usually 
better constrained and less reliant on atmospheric
models than effective temperature and radius, which can substantially vary with 
assumptions about cloud properties, chemical composition,
and sources of opacity.  
Below are summaries of the major assumptions 
(in roughly descending order) 
involved in the inference of planet masses using atmospheric and evolutionary models along
with notable advantages, drawbacks, and limitations of various techniques.

\begin{itemize}
\item \textbf{Initial conditions and formation pathway.}   
The most important assumption 
is the 
amount of initial energy and entropy a planet begins with following its formation.
This defines its evolutionary pathway, which is embodied in three broad classes informed by 
formation mechanisms.  

Hot-start models begin with arbitrarily large radii and oversimplified, idealized initial conditions 
that generally ignore the effects of accretion and mass assembly.  
As such, they represent the most luminous outcome and correspond to the most optimistically
(albeit unrealistically) low mass estimates.
Ironically, hot-start grids are nearly unanimously adopted for estimating the masses 
of young brown dwarfs and giant planets 
even though the early evolution in these models is the least reliable.
The most widely used hot-start models for imaged planets are
the Cond and Dusty grids from \citet{Baraffe:2003bj} and \citet{Chabrier:2001bf},
\citet{Burrows:1997jq}, and \citet{Saumon:2008im}.

Cold-start models were made prominent by \citet{Marley:2007bf} and \citet{Fortney:2008ez} in the context of direct imaging
as an attempt to emulate a more realistic formation scenario for giant planets through core accretion.
In this model, accretion shocks radiate the gravitational potential energy of infalling gas as a giant planet grows.  
After formation, these planets begin cooling with much lower luminosities and 
initial entropies compared to the hot-start scenario, taking between $\sim$10$^8$ and $\sim$10$^9$ years to 
converge with hot-start cooling models depending on the planet mass.  
The observational implications of this are severe: planets formed from core accretion may be orders of magnitude
less luminous than those produced from cloud fragmentation or disk instability.  While this
may offer a diagnostic for the formation route if the mass of a planet is independently measured, 
it also introduces considerable uncertainty in the more typical case when only an age and luminosity are known.
For example, 51~Eri~b may be as low as 2~\Mjup \ or as high as 12~\Mjup \ depending on which cooling model 
(hot or cold)
is assumed (\citealt{Macintosh:2015ewa}).

This picture is made even more complicated by large uncertainties in the details of cold-start models. 
The treatment of accretion shocks, circumplanetary disks, core mass, and even deuterium
burning for the most massive planets can dramatically influence the initial entropy and luminosity evolution
of planets (\citealt{Mordasini:2012jy}; \citealt{Bodenheimer:2013ki}; \citealt{Mordasini:2013cr}; \citealt{Owen:2016bu}).  
This motivated a class of warm-start models with intermediate initial entropies that probably better 
reflect dissipative accretion shocks that occur in nature (\citealt{Spiegel:2012ea}; \citealt{Marleau:2013bh}).  
Unfortunately, the relevant details of giant planet assembly 
are poorly constrained by observations.  There is also likely to be intrinsic scatter in the 
initial conditions for a given planet which may result in large degeneracies in the planet mass, core mass,
and accretion history for young gas giants with the same age and luminosity.  It is quite possible, for instance,
that the HR 8799 c, d, and e planets which all share the same age and nearly the same luminosity could have 
very different masses.

\begin{figure}
  \vskip -.4 in
  \hskip -1.5 in
  \resizebox{6.9in}{!}{\includegraphics{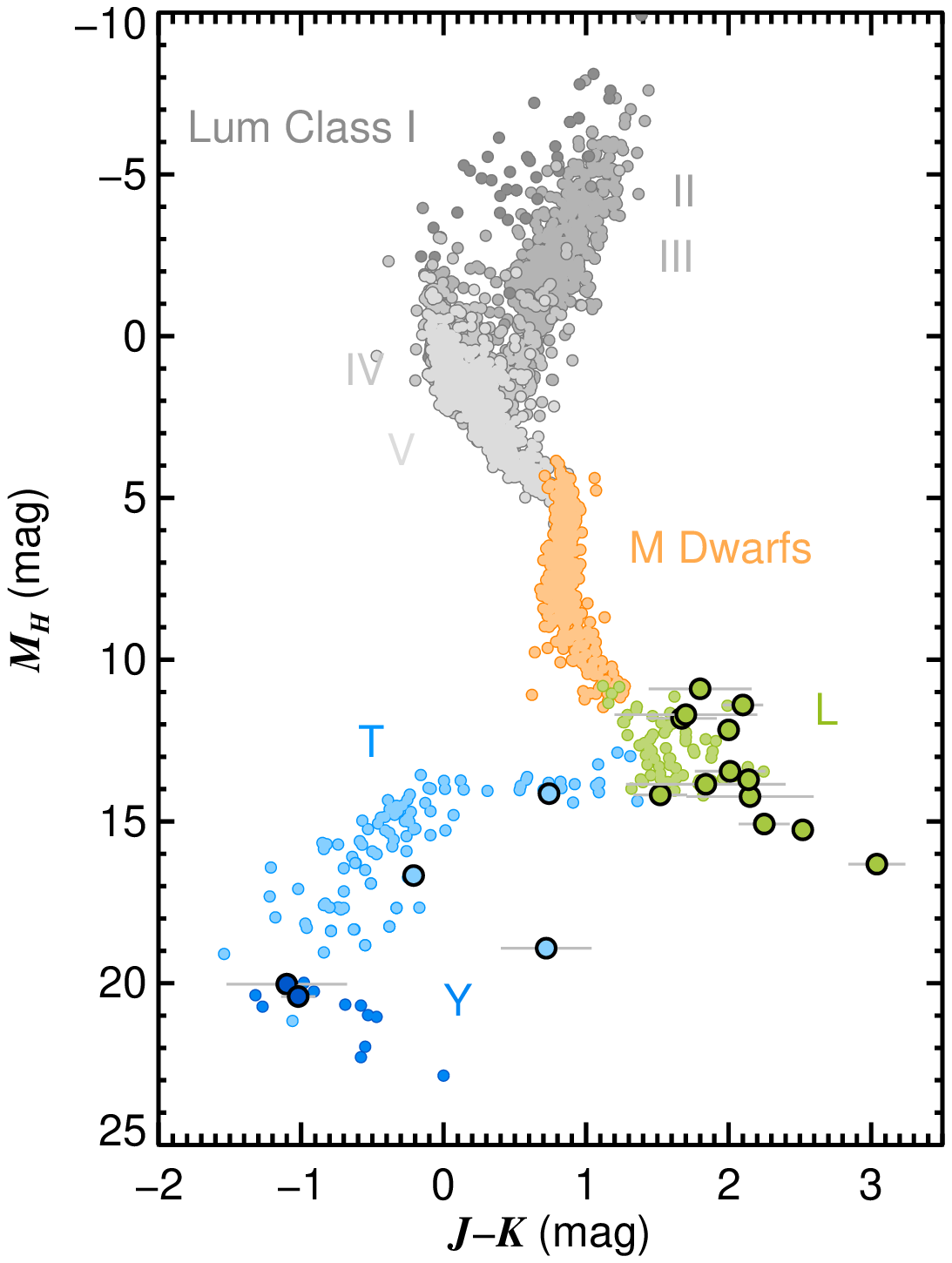}}
  \vskip -0.05 in
  \caption{The modern color-magnitude diagram spans nearly 35 magnitudes in
  the near-infrared and 5 magnitudes in $J$--$K$ color.  
  The directly imaged planets (bold circles) extend the L dwarf sequence 
  to redder colors and fainter absolute 
  magnitudes owing to a delayed transition from cloudy atmospheres to 
  condensate-free T dwarfs at low surface gravities.  The details of this
  transition for giant planets remains elusive.
  OBAFGK stars (gray) are from the extended $Hipparcos$ compilation 
  XHIP (\citealt{Anderson:2012cu}); 
  M dwarfs (orange) are 
  from \citet{Winters:2015ji};
  late-M dwarfs (orange), L dwarfs (green), and T dwarfs (light blue) are from \citet{Dupuy:2012bp}; 
  and Y dwarfs (blue) are compiled largely from 
  \citet{Dupuy:2014iz}, \citet{Tinney:2014bl}, and \citet{Beichman:2014jr} by 
  T. Dupuy (2016, private communication).
   Directly imaged planets or planet candidates (bold circles) represent all companions from 
   Table~\ref{tab:planets} with near-infrared photometry
   and parallactic distances.
  \label{fig:cmd} } 
  \vskip -.1 in
\end{figure}

\item \textbf{Stellar age.}
After bolometric luminosity, which is generally uncomplicated to estimate for imaged planets, 
the age of the host star is the most sensitive parameter on which the mass of an imaged companion depends. 
It is also one of the most difficult quantities to accurately determine and usually relies on stellar evolutionary models
or empirical calibrations.  Recent reviews on this topic include \citet{Soderblom:2010kr}, \citet{Jeffries:2014js}, 
and \citet{Soderblom:2014ve}.   
Figure~\ref{fig:lum_age} shows the current census of imaged companions near and below
the deuterium-burning limit with both age and luminosity measurements (from Table~\ref{tab:planets}).
Apart from uncertainties in the formation history of these objects, age uncertainties
dominate the error budget for inferring masses.

Clusters of coeval stars spanning a wide range of masses 
provide some of the best age constraints but are still dominated by systematic errors.  
Several star-forming regions and young moving groups in particular have been systematically 
adjusted to older ages over the past few years, which has propagated to the ages and masses of planets 
in those associations (\citealt{Pecaut:2012gp}; \citealt{Binks:2014gd}; \citealt{Kraus:2014ur}; \citealt{Bell:2015gw}).
The implied hot-start mass for $\beta$~Pic~b, for example, increases by several Jupiter masses 
(corresponding to several tens of percent) assuming the planet's age is $\approx$23~Myr instead of $\approx$12~Myr (\citealt{Mamajek:2014bf}; although see the next bullet point).

For young field stars, distant stellar companions can help age-date the entire system.  For example, 
the age of Fomalhaut was
recently revised to $\sim$400~Myr from $\sim$200~Myr in part due to constraints from its wide 
M dwarf companions (\citealt{Mamajek:2012ga}; \citealt{Mamajek:2013gzb}).
Ultimately, if the age of a host star is unknown, the significance and interpretation of a faint companion is limited if 
basic physical properties like its mass are poorly constrained.

\item \textbf{Epoch of planet formation.} 
Planets take time to form so they are not exactly coeval with their host stars.  Their ages may span
the stellar age to the stellar age minus $\sim$10~Myr depending on the timescale for giant planets to assemble.  
Planets formed via cloud fragmentation or disk instability might be nearly coeval with their host star, but those formed by
core accretion are expected to build mass over several Myr. While this difference is negligible
at intermediate and old ages beyond a few tens of Myrs, it can have a large
impact on the inferred masses of the youngest planets ($\lesssim$20~Myr).
For example, if the age of the young planetary-mass companion 2M1207--3932~b is assumed to be coeval with the TW~Hyrdae Association ($\tau$ = 10~$\pm$~3~Myr) then
its hot-start mass is $\approx$5~\Mjup.  On the other hand, if its formation was delayed by 8~Myr ($\tau$ = 2 Myr) then its mass is only $\approx$2.5~\Mjup.

\begin{figure*}
  \vskip -.6 in
  \hskip -0 in
  \resizebox{7.6in}{!}{\includegraphics{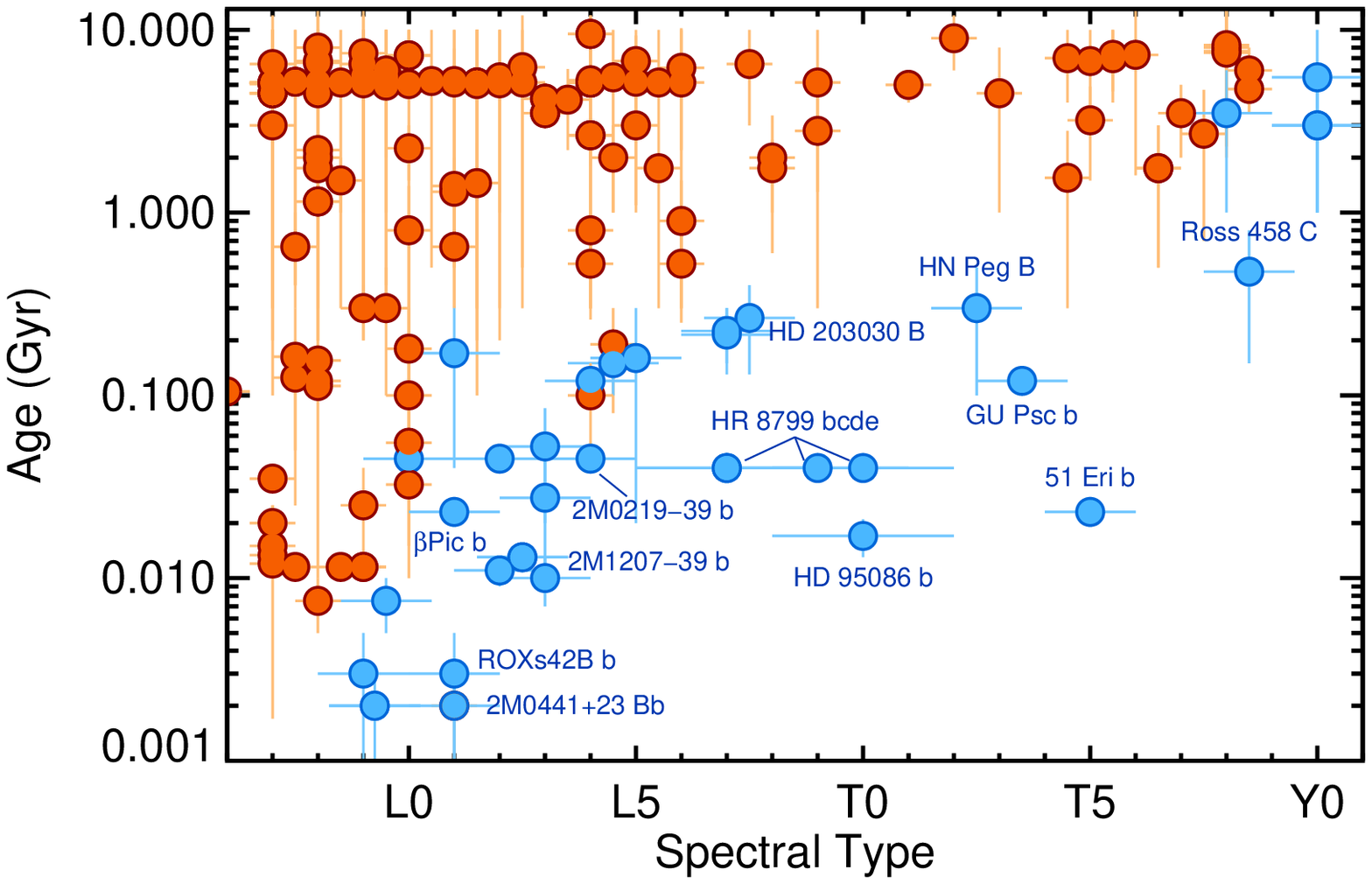}}
  \vskip -1.15 in
  \caption{Ultracool substellar companions with well-constrained ages and spectroscopically-derived classifications.  Red circles are low-mass stars and brown dwarfs ($>$13~\Mjup) while blue circles show companions near and below the deuterium-burning limit.  Companions are primarily from \citet{Deacon:2014ey} together with additional discoveries from the literature.  \label{fig:spt_age} } 
\end{figure*}

\item \textbf{Atmospheric models.}  
Atmospheric models can influence the inferred masses of imaged exoplanets in several ways.
They act as surface boundary conditions for evolutionary models and regulate radiative cooling through molecular and
continuum opacity sources.  This in turn impacts the luminosity evolution of giant planets, albeit minimally 
because of the weak dependence on mean opacity ($L(t)$ $\propto$ $\kappa^{0.35}$; \citealt{Burrows:1993kt}; \citealt{Burrows:2001wq}).  
Even the unrealistic cases of permanently dusty and perpetually condensate-free photospheres do not dramatically affect the luminosity evolution of 
cooling models or mass determinations using age and bolometric luminosity (\citealt{Baraffe:2002di}; \citealt{Saumon:2008im}), although more realistic (``hybrid'') models accounting for the evolution and dissipation of clouds at the L/T transition can influence the shape of cooling curves in slight but significant ways (\citealt{Saumon:2008im}; \citealt{Dupuy:2015gl}).

On the other hand, mass determinations in color-magnitude space are highly sensitive to atmospheric models and can
result in changes of several tens of percent depending on the specific treatment of atmospheric condensates.
Dust reddens spectra and can modify the near-infrared colors and absolute magnitudes of ultracool objects  
by several magnitudes.  This introduces another source of uncertainty if the spectral shape is poorly constrained, though
the difference between dusty and cloud-free models is smaller at longer wavelengths and higher temperatures.  

One of the most important and unexpected empirical results to emerge from direct imaging 
has been the realization that young brown dwarfs and massive
planets retain photospheric clouds even at low effective temperatures where older, high-gravity brown dwarfs 
have already transitioned to T dwarfs (\citealt{Metchev:2006bq}; \citealt{Chauvin:2004cy}; \citealt{Marois:2008ei}; 
\citealt{Patience:2010hf}; \citealt{Bowler:2010ft}; \citealt{Faherty:2012bc}; \citealt{Bowler:2013ek}; \citealt{Liu:2013gya}; \citealt{Filippazzo:2015dv}).
This is demonstrated in Figure~\ref{fig:cmd}, which shows the location of imaged companions near and below the deuterium-burning limit on the near-infrared color-color diagram.  At young ages, warm giant planets are significantly redder than the field population of brown dwarfs, and several of the most extreme examples have anomalously low absolute magnitudes.
For old brown dwarfs, this evolution from dusty, CO-bearing L dwarfs to cloud-free, methane-dominated T dwarfs takes place over a narrow 
temperature range ($\sim$1200--1400~K)  but occurs at a lower (albeit still poorly constrained) temperature
for young gas giants.  The lack of methane is likely caused by disequilibrium carbon chemistry 
at low surface gravities as a result of vigorous vertical mixing (e.g., \citealt{Barman:2011fe}; \citealt{Zahnle:2014hl}; \citealt{Ingraham:2014gx}; \citealt{Skemer:2014hy}), while the 
preservation of photospheric condensates can be explained by a dependency of cloud base pressure 
and particle size on surface gravity (\citealt{Marley:2012fo}).
Unfortunately, the dearth of known planets between $\sim$L5--T5 is the main limitation to understanding this transition in detail (Figure~\ref{fig:spt_age}).

In principle, the mass of a planet can also be inferred by fitting synthetic spectra to the planet's 
observed spectrum or multi-band photometry.   The mass can then be obtained from best-fitting model as follows:

\begin{equation}
M_p \ (M_\mathrm{Jup}) = 12.76 \times 10^{\log(g) - 4.5~\mathrm{dex}}  \left( \frac{R}{R_\mathrm{Jup}} \right)^2.
\end{equation}

\noindent Here $\log(g)$ is the surface gravity (in cm s$^{-2}$) and $R$ is the 
planet's radius.  The radius can either be taken from evolutionary models or alternatively from 
the multiplicative factor that scales the emergent model spectrum to the observed flux-calibrated 
spectrum (or photometry) of the planet.
This scale factor corresponds to the planet's radius over
its distance, squared ($R^2$/$d^2$; see \citealt{Cushing:2008kb} for details).

Clearly the inferred mass is very sensitive to both the surface gravity and the radius.  
In practice, gravity is usually poorly constrained for model fits to brown dwarf and giant planet spectra 
because its influence on the emergent spectrum is more subtle 
(e.g., \citealt{Cushing:2008kb}; \citealt{Bowler:2011gw}; \citealt{Barman:2011fe}; \citealt{Macintosh:2015ewa}).
In addition, the scale factor strongly depends on the model effective temperature ($\propto$ $T_\mathrm{eff}^{-4}$), which is
typically not known to better than $\sim$100~K.
Altogether, the current level of systematic imperfections present in atmospheric models and observed spectra of exoplanets (e.g., \citealt{Greco:2016ww}) 
mean that masses cannot yet be reliably measured from fitting grids of synthetic spectra.

\item \textbf{Deuterium burning history.}  
As brown dwarfs with masses between about 13~\Mjup \ and 75~\Mjup \ contract, their core temperatures become 
hot enough to burn deuterium, though not at sufficient rates to balance surface radiative losses 
(e.g., \citealt{Kumar:1963ht}; \citealt{Burrows:1993kt}).\footnote{ Solar-metallicity brown dwarfs with masses above $\approx$63~\Mjup \ can also burn lithium.  This limit changes slightly for non-solar values (\citealt{Burrows:2001wq}).}  
The onset and timescale of deuterium burning varies primarily with mass but also with metallicity, helium fraction, 
and initial entropy (e.g., \citealt{Spiegel:2011ip}); lower-mass brown dwarfs
take longer to initiate deuterium burning than objects near the hydrogen-burning minimum mass.
This additional transient energy source delays the otherwise invariable cooling and causes 
luminosity tracks to overlap.  Thus, objects with the same luminosity and
age can differ in mass depending on their deuterium-burning history.  
Many substellar companions fall in this ambiguous region, complicating mass determinations by up to a factor of $\sim$2 (Figure~\ref{fig:lum_age} and Table~\ref{tab:planets}).  
With a large sample of objects in this region, spectroscopy may ultimately be able to distinguish higher- and lower-mass 
scenarios through relative surface gravity measurements (\citealt{Bowler:2013ek}). 

\item \textbf{Planet composition.}  
The gas and ice giants in the Solar System are enriched in heavy elements compared to solar values.
The specific mechanism for this enhancement is still under debate but exoplanets formed via core
accretion are expected to show similar compositional and abundance ratio differences compared to
their host stars, whereas planets formed through cloud fragmentation or disk instability are probably quite similar
to the stars they orbit.
The bulk composition of planets modifies their atmospheric opacities and influences both their emergent spectra
and luminosity evolution (\citealt{Fortney:2008ez}).  
A common practice when deriving masses for imaged planets 
is to assume solar abundances, which is largely dictated by the availability of published  
atmospheric and evolutionary models.
Many of these assumptions can be removed with atmospheric retrieval methods by directly fitting for
atomic and molecular abundances (\citealt{Lee:2013br}; \citealt{Line:2014eu}; \citealt{Todorov:2015vha}).

\item \textbf{Additional sources of uncertainty.}  
A number of other factors and implicit assumptions can also introduce 
random and systematic uncertainties in mass derivations.
Different methods of PSF subtraction can bias photometry if planet self-subtraction or speckle over-subtraction is not
properly corrected (e.g., \citealt{Marois:2006df}; \citealt{Lafreniere:2007bg}; \citealt{Soummer:2012ig}).  
 Photometric variability from rapidly changing or 
rotationally-modulated surface features can introduce 
uncertainties in relative photometry (e.g., \citealt{Radigan:2014dj}; 
\citealt{Metchev:2015dr}; \citealt{Zhou:2016gc}; \citealt{Biller:2015hza}).  The host star can also be variable if
it has unusually large starspot coverage or if it is very young and happens to 
have an edge-on disk (e.g., TWA~30A; \citealt{Looper:2010cb}).

Multiplicity can also bias luminosity measurements.  Roughly 20\% of brown dwarfs are 
close binaries with separation distributions peaking near 4.5~AU and mass ratios approaching 
unity (e.g., \citealt{Burgasser:2007uf}; \citealt{Kraus:2012et}; \citealt{Duchene:2013il}).  
If some planetary-mass companions form in the same
manner as brown dwarfs, and if the same trends in multiplicity continue into the planetary regime,
then a small fraction of planetary-mass companions are probably close, unresolved, equal-mass binaries.
These systems will appear twice as luminous.

If atmospheric chemistry or cloud structure varies latitudinally  then orientation and viewing angle 
could be important. 
For the youngest protoplanets embedded in their host stars' circumstellar disks, 
accretion streams might dominate over thermal photospheric emission, complicating luminosity
measurements and mass estimates (e.g., LkCa15~b and HD~100546~b; \citealt{Kraus:2012gk}; \citealt{Quanz:2013ii}). 
Approaches to applying bolometric corrections, measuring 
partly opaque coronagraphs and neutral density filters, 
finely interpolating atmospheric and
evolutionary model grids, or converting models between filter systems 
(for example, $CH_4S$ to $K$) may vary.
Finally, additional energy sources like radioactivity or stellar insolation are assumed to be negligible
but could impact the luminosity evolution of some exoplanets.

\end{itemize}

\subsection{Measuring Masses}

No imaged planet has yet had its mass measured.  The most robust, model-independent way to do so
is through dynamical interactions with other objects.
Because planets follow mass-luminosity-age relationships, knowledge of all three parameters
are needed to test cooling models.  Once a mass is measured, its age (from the host star) and
bolometric luminosity (from its distance and spectral energy distribution) enable precision model tests,
although an assumption about energy losses from accretion via hot-, warm-, or cold-start 
must be made.
Nevertheless, if all hot-start models overpredict the luminosities of giant planets,
that would suggest that accretion history is indeed an important factor in 
both planet formation and realistic cooling models.  
Below is a summary of methods to measure substellar masses.

\begin{itemize}
\item \textbf{Dynamical masses.}  
Most close-in ($<$100~AU) planets have shown significant orbital motion since their discoveries (Table~\ref{tab:planets}).
This \emph{relative} motion provides a measure of the total mass of the system ($M_\mathrm{star}$+$M_\mathrm{planet}$).  
If stationary background stars can simultaneously be observed with the planet-star pair
then \emph{absolute} astrometry is possible.  This then gives individual masses
for each component ($M_\mathrm{star}$ and $M_\mathrm{planet}$ separately).
Unfortunately the long orbital periods and lack of nearby background stars for the present census of imaged planets means
this method is currently impractical to measure masses. 

Relative astrometry can also be combined with radial velocities to measure a planet's mass.
Assuming the visual orbit and total mass are well constrained from imaging, the mass of 
the companion can be measured by monitoring the line of sight reflex motion of the host star (e.g., \citealt{Crepp:2012eg}).
This treats the system as a single-lined spectroscopic binary, giving the mass function 
$m_\mathrm{p}^3 \sin^3 i$/$M_\mathrm{tot}^2$,
where $M_\mathrm{tot}$ is the measured total mass, $i$ is the measured inclination, and $m_\mathrm{p}$
is the mass of the planet.
If precise radial velocities are not possible for the host star because it has an early spectral type (with
few absorption lines) or
high levels of stellar activity (RV jitter)
then RV monitoring of the planet can also yield its mass.
This can be achieved by combining adaptive optics imaging and high-resolution near-infrared 
spectroscopy to spatially separate the star and planet, as has been demonstrated with 
$\beta$~Pic~b (\citealt{Snellen:2014kz}).
Soon $Gaia$ will produce precise astrometric measurements of the host stars of imaged planets.
Together with orbit monitoring though high-contrast imaging, this may offer another way to directly
constrain the masses of imaged planets. 

Close substellar binaries offer another approach.  Their orbital periods are typically faster
and, in rare cases when such binaries themselves orbit a star, 
the age of the tertiary components can be adopted from the host star.  
Several brown dwarf-brown dwarf masses have been measured in this fashion:
HD~130948~BC (\citealt{Dupuy:2009jq}), Gl~417~BC (\citealt{Dupuy:2014iz}),
and preliminary masses for $\epsilon$~Indi~Bab (\citealt{Cardoso:2009jf}).
Isolated substellar pairs are also useful for dynamical mass measurements but their
ages are generally poorly constrained unless they are members of young clusters
or moving groups.
No binaries with both components unambiguously residing in the planetary-mass regime are known, but there
is at least one candidate (WISE~J014656.66+423410.0; \citealt{Dupuy:2015dza}).  

\item \textbf{Keplerian disk rotation.}  
Dynamical masses for young protoplanets may eventually be possible using ALMA through
Keplerian rotation of circumplanetary disks.
This requires resolving faint gas emission lines (e.g., CO~$J$=3--2 or CO~$J$=2--1) from the planet 
both spatially and spectrally, something
that has yet to be achieved for known young planets harboring 
subdisks (e.g., \citealt{Isella:2014fz}; \citealt{Bowler:2015hx}).
Although challenging, this type of measurement can act as a detailed probe 
of the initial conditions of giant planet formation and evolution.

\item \textbf{Stability analysis.}  
Numerical modeling of planets and their interactions with debris disks, protoplanetary disks, or additional 
planets offers another way to constrain the mass of an imaged planet.  If their masses are too low, planets will not
be able to gravitationally shape dust and planetesimals in a manner consistent with observations.  
Modeling of the disks and companions orbiting Fomalhaut and $\beta$~Pic illustrate this approach; 
independent constraints on the orbit and masses of these companions can be made  
by combining spatially-resolved disk structures (a truncated, offset dust ring encircling 
Fomalhaut and a warped inner
disk surrounding $\beta$~Pic~b) and observed orbital motion
(e.g., \citealt{Chiang:2009co}; \citealt{Dawson:2011eu}; \citealt{Kalas:2013hpa}; \citealt{Beust:2014dj}; 
\citealt{MillarBlanchaer:2015ha}). 
Likewise, if a planet's mass is too high then
it carves a larger disk gap or may destabilize other planets in the system through mutual interactions.  
For example, detailed $N$-body simulations of HR~8799's planets have shown that they 
must have masses $\lesssim$10--20~\Mjup \ --- consistent with giant planet evolutionary models --- or they would have become dynamically unstable 
by the age of the host star 
(e.g., \citealt{Gozdziewski:2009fv}; \citealt{Fabrycky:2010fi}; \citealt{Currie:2011iz}; \citealt{Sudol:2012gn}; \citealt{Gozdziewski:2014gz}).

\item \textbf{Disk morphology.}  
Large-scale structures in disks --- clumps, asymmetries, warps, gaps, rings, truncated edges, spiral arms, 
and geometric offsets --- can
also be used to indirectly infer the presence of unseen planets and predict their masses and locations
(e.g., \citealt{Wyatt:1999ty}; \citealt{Ozernoy:2000dd}; \citealt{Kenyon:2004wr}).  
This approach relies on assumptions about disk surface density
profiles and grain properties, so it is not a completely model-free measurement, but it is potentially sensitive to
planet masses as low as a few tens of Earth masses (\citealt{Rosotti:2016vsa}). 
It also enables an immediate mass evaluation without the need for long-term orbit monitoring.
Recently, \citet{Dong:2015fg} and \citet{Zhu:2015fa} presented a novel approach along these lines 
to predict the locations and masses 
widely-separated companions inducing spiral arms on a circumstellar disk 
(see also \citealt{Dong:2016kc}, \citealt{Dong:2016wc}, and \citealt{Jilkova:2015joa}).  This may prove to be
a valuable way to constrain masses of planetary companions at extremely wide orbital distances.

\end{itemize}

\begin{figure*}
  \vskip -.4 in
  \hskip -1. in
  \resizebox{9.in}{!}{\includegraphics{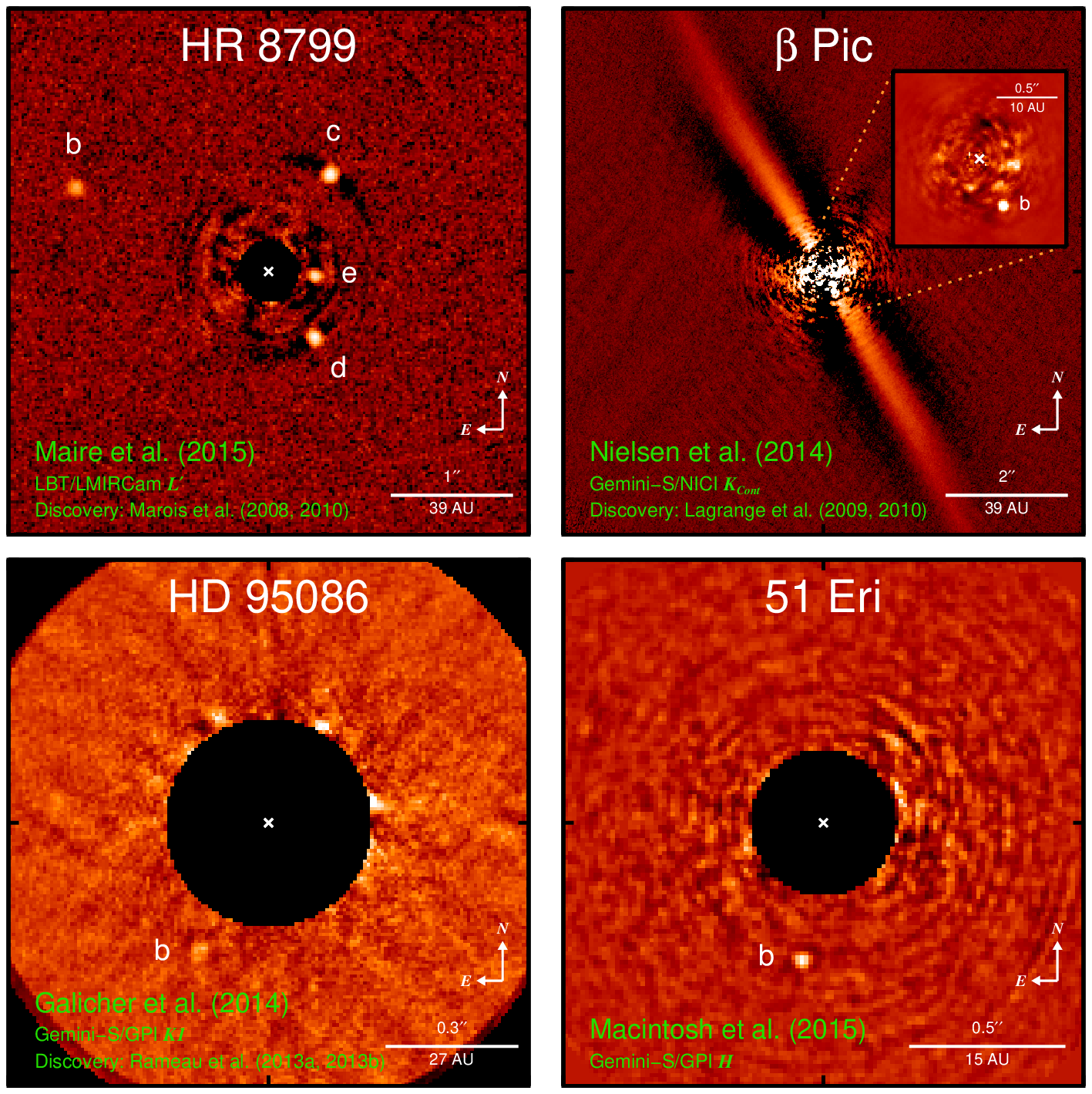}}
  \vskip -0.3 in
  \caption{Gallery of imaged planets at small separations ($<$100~AU).  HR~8799 harbors four massive planets (5--10~\Mjup) at orbital distances of 15--70~AU (\citealt{Marois:2008ei}; \citealt{Marois:2010gpa}),  $\beta$~Pic
  hosts a nearly edge-on debris disk and a $\approx$13~\Mjup \ planet at 9~AU (\citealt{Lagrange:2009hq}; \citealt{Lagrange:2010fsa}),  a $\approx$5~\Mjup \ planet orbits HD~95086 at 56~AU 
  (\citealt{Rameau:2013vh}; \citealt{Rameau:2013ds}), and 51~Eri hosts a $\sim$2~\Mjup \ planet 
  at 13~AU (\citealt{Macintosh:2015ewa}).  Images are from \citet{Maire:2015ek}, 
  \citet{Nielsen:2014js}, \citet{Galicher:2014er}, and \citet{Macintosh:2015ewa}. \label{fig:planetgallery} } 
\end{figure*}

\section{Survey of Surveys}{\label{sec:surveys}}

Myriad large high-contrast imaging surveys have been carried out over the past decade\footnote{The basic properties 
for most of these programs until 2014 are summarized in Table 1 of  \citet{Chauvin:2015jy}.}.  The most impactful
programs are highly focused, carefully designed with well-understood biases, and have meticulously-selected target lists 
to address specific science questions.
The advantages of large surveys include homogeneous observations, instrument setups, 
data reduction pipelines, and statistical treatments of the results. 

\begin{figure*}
  \vskip -1 in
  \hskip -1.2 in
  \resizebox{9.in}{!}{\includegraphics{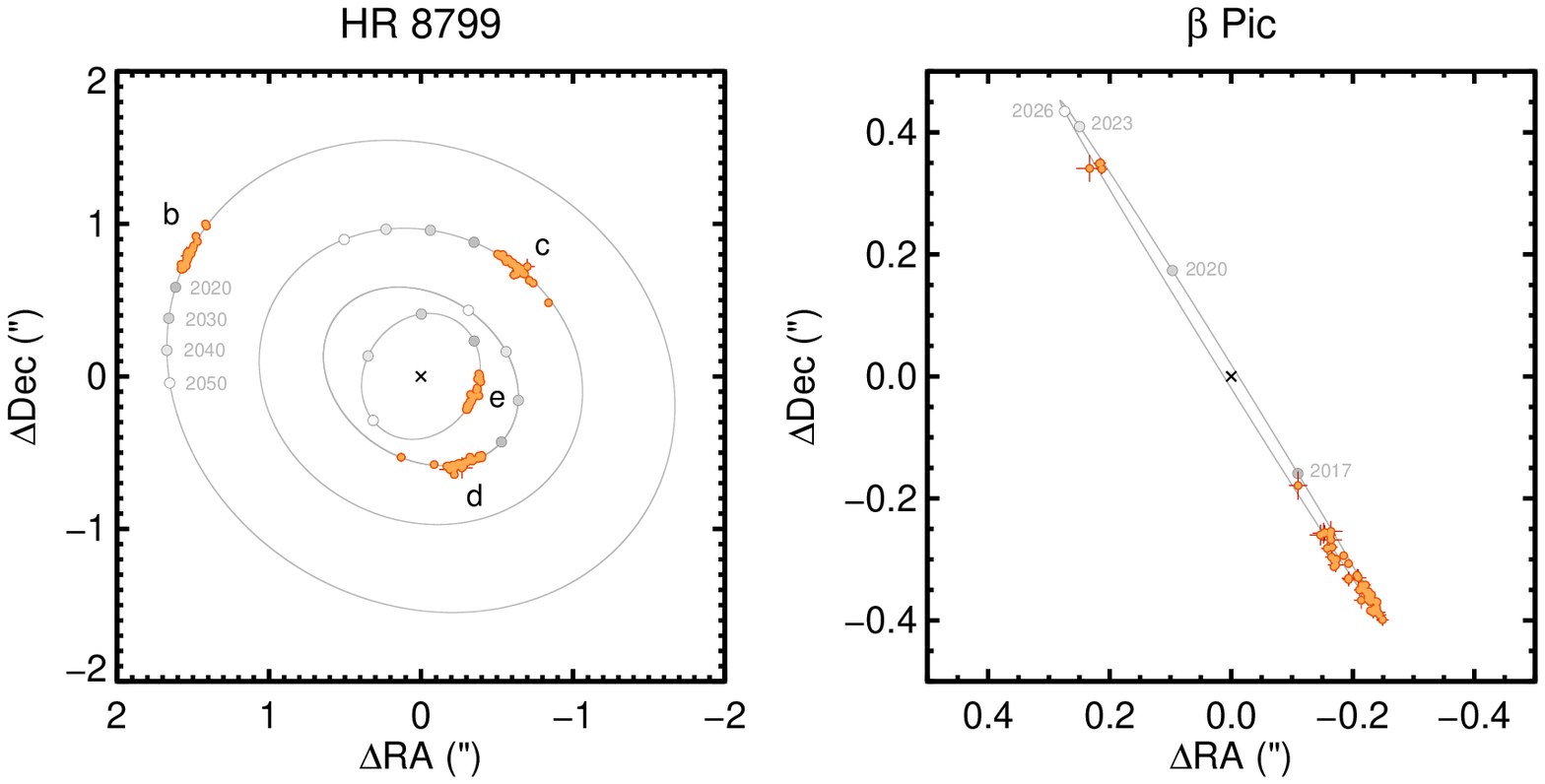}}
  \vskip -2.1 in
  \caption{The orbits of HR~8799~bcde and $\beta$~Pic~b.  Astrometric monitoring has revealed significant orbital motion over the past decade, enabling measurements of their Keplerian orbital parameters and offering clues about their dynamical history.  The HR~8799 planets are on low-eccentricity orbits with some evidence that HR~8799~d may be mutually misaligned with the other planets, which otherwise appear to be coplanar (e.g., \citealt{Pueyo:2015cx}).  $\beta$~Pic~b follows a nearly edge-on orbit with a low eccentricity (e.g., \citealt{MillarBlanchaer:2015ha}).  Astrometric measurements are compiled in Appendix~\ref{tab:astrometry}. Orbits depicted here are from \citet{Zurlo:2016hl} and \citet{Nielsen:2014js}.     \label{fig:astrometry} } 
\end{figure*}

Below are summaries of the most substantial 
high-contrast imaging surveys carried out to date with a focus on deep adaptive optics imaging programs
that routinely reach planetary masses and employ modern observing and post-processing techniques 
to suppress speckle noise.  
These surveys produced the first wave of discoveries (Figure~\ref{fig:planetgallery}), opening 
the door to directly characterizing the atmospheres of exoplanets as well as their orbits 
through astrometric monitoring (Figure~\ref{fig:astrometry} and Appendix~\ref{tab:astrometry})
This section follows
an historical approach by outlining early ground- and space-based experiments, the first generation of planet-finding instruments and associated surveys,
and the next generation of instruments characterized by extreme adaptive optics systems with exceptionally high Strehl ratios.

\subsection{Early Surveys}

Early high-contrast imaging surveys in search of closely-separated brown dwarf companions 
and giant planets were conducted with
speckle interferometry (\citealt{Henry:1990ig}),
image stabilizers (\citealt{Nakajima:1994ea}),
$HST$ (\citealt{Sartoretti:1998uc}; \citealt{Schroeder:2000tk}; \citealt{Brandner:2000vp}; \citealt{Lowrance:2005ci}; \citealt{Luhman:2005cn}),
speckle cameras (\citealt{Neuhauser:2003tb}), 
or newly-commissioned adaptive optics systems from the ground with facility instruments
(\citealt{Oppenheimer:2001kl}; \citealt{Macintosh:2001tr}; \citealt{Chauvin:2003gi}; 
\citealt{McCarthy:2004hl}; \citealt{Carson:2005ia}; \citealt{Nakajima:2005cm}; \citealt{Tanner:2007hu}).
When PSF subtraction was performed, it usually entailed roll-subtraction
(for $HST$; e.g., \citealt{Liu:2004kk}), self-subtraction with a rotated PSF, or reference star subtraction.

Some of these pioneering programs are especially noteworthy for their depth and emphasis on statistical results.
\citet{Lowrance:2005ci} targeted 45 single young A--M stars with NICMOS in the $F160W$ 
filter (1.6~$\mu$m) on board $HST$.  Two brown dwarfs were uncovered, TWA~5~B (\citealt{Lowrance:1999ck}) and
HR~7372~B (\citealt{Lowrance:2000ic}), as well as Gl~577~BC, a tight binary companion near the 
hydrogen-burning limit.
\citet{Masciadri:2005gl} used NaCo at the VLT to obtain deep adaptive optics imaging of 28 young nearby stars.
No substellar companions were found, but the importance of thoroughly reporting survey results is highlighted,
even for non-detections,
a theme that continues today.
Focusing exclusively on young moving group members in $L'$-band enabled \citet{Kasper:2007dm} to reach
exceptionally low limiting masses for a sample of 22 stars with NaCo.
The Palomar and Keck adaptive optics survey by \citet{Metchev:2009ky} is another especially
valuable contribution; they imaged 266 FGK stars and discovered two brown dwarf companions,
HD~49197~B (\citealt{Metchev:2004kl}) and HD~203030~B (\citealt{Metchev:2006bq}),
implying a substellar occurrence rate of 3.2$^{+3.1}_{-2.7}$\%.  HD~203030~B was 
the first young brown dwarf for which signs of a 
discrepancy between the field spectral type-effective temperature sequence was recognized, now understood
as a retention of clouds to lower effective temperatures at low surface gravities.
These groundbreaking surveys helped define the scientific motivation, 
framework, and early expectations for the first generation planet-finding instruments and 
larger observing programs.

\subsection{The First Generation: Dedicated Instruments, Expansive Surveys, and Innovative Speckle Suppression Techniques}{\label{sec:firstgen}}

High-contrast imaging is largely driven by advances in instrumentation and speckle suppression.
The first wave of instruments specifically designed to image giant planets
gave rise to large surveys ($N$$\approx$50--500) targeting mostly young nearby stars.
Deep observations in pupil-tracking mode (angular differential imaging) have become standardized as a way to 
distinguish quasi-static speckles from planets (\citealt{Marois:2006df}).
This era is also characterized by the advent of advanced PSF subtraction techniques
to optimally remove speckles during post-processing.  Two especially important algorithms are the Locally Optimized Combination of Images
(LOCI; \citealt{Lafreniere:2007bg}), which is based on least-squares minimization of residual speckle noise, and
Karhunen-Lo{\`e}ve Image Projection (KLIP; \citealt{Soummer:2012ig}), a computationally-fast method based on principal component analysis.
The introduction of these new methods gave rise to an array of sophisticated data reduction pipelines with additional features 
aimed at minimizing biases and avoiding both self- and over-subtraction of planet flux in ADI and SDI datasets
(\citealt{Marois:2010hs}; \citealt{Amara:2012hv}; \citealt{Pueyo:2012ft}; \citealt{Meshkat:2013ej}; 
\citealt{Wahhaj:2013fq}; \citealt{Brandt:2013in}; \citealt{Fergus:2014hv}; \citealt{Mawet:2014ga};  \citealt{Marois:2014ep}; 
 \citealt{Currie:2014fm}; \citealt{Cantalloube:2015km}; \citealt{Rameau:2015fg}; \citealt{Wahhaj:2015jz}; \citealt{Savransky:2015kg}; \citealt{Dou:2015gk}; 
\citealt{Hagelberg:2016kh}; \citealt{Gonzalez:2016ul}).  

The suite of instrumentation for high-contrast imaging has ballooned over the past 15 years and
includes dual-channel imagers, infrared wavefront sensors, 
non-redundant aperture masking interferometry, adaptive secondary mirrors,
integral field units, high-order adaptive optics systems, and 
specialized coronagraphs (e.g., apodized Lyot coronagraph, annular groove phase mask coronagraph, 
vector vortex coronagraph, apodizing phase plate, 
and four quadrant phase mask; \citealt{Rouan:2000jn}; \citealt{Guyon:2005dp}; \citealt{Soummer:2005co}; 
\citealt{Mawet:2005vb}; \citealt{Kenworthy:2007vz}; \citealt{Mawet:2010el}).  
Many of these have been implemented
in the first generation of instruments in part as testbeds for regular use in second-generation systems.
These instruments are reviewed in detail in  \citet{Guyon:2006jp}, \citet{Beuzit:2007tj}, \citet{Oppenheimer:2009gh}, 
\citet{Perryman:2011uo}, and \citet{Mawet:2012il}.

\subsubsection{VLT and MMT Simultaneous Differential Imager Survey}{\label{sec:gdps}}

This survey (PI: B Biller) targeted 45 young stars between 2003--2006 with ages $\lesssim$250~Myr and distances within 50~pc 
using Simultaneous Differential Imagers mounted on the VLT and MMT (\citealt{Biller:2007ht}).
It was among the first to utilize simultaneous differential imaging to search for cold planets
around a large sample of young stars.  The SDI method takes advantage
of expected spectral differences between the star, which has a nearly flat continuum, 
and cool, methanated planets by simultaneously imaging in multiple narrow-band filters
across this deep absorption feature at 1.6~$\mu$m.  Because speckles radially scale with wavelength
while real objects remain stationary, their observations also had some sensitivity to warmer planets without
methane (though it is now clear that the onset of methane occurs at lower temperatures for giant planets than for
brown dwarfs).  No substellar companions were found, which ruled out a
linearly-flat extension of close-in giant planets out to 45~AU with high confidence.

\subsubsection{GDPS: Gemini Deep Planet Survey}{\label{sec:gdps}}

GDPS (PI: D. Lafreni\`{e}re) was a large high-contrast imaging 
program at the Gemini-North 8.1-m telescope with the NIRI camera and Altair AO system 
focusing on 85 stars, 16 of which were identified as close multiples (\citealt{Lafreniere:2007cv}).  
The sample contained a mix of nearby GKM stars within 35~pc comprising then-known or suspected nearby young moving group members,
stars with statistically young ages, and several others harboring circumstellar disks.
Altogether the ages span 10~Myr to $\sim$5~Gyr.
The observations were taken in ADI mode with the $CH_4S$ filter, and PSF subtraction was
carried out with the LOCI algorithm (\citealt{Lafreniere:2007bg}).
No substellar companions were discovered, implying an occurrence rate of
$<$23\% for $>$2~\Mjup \ planets between 25--420~AU and 
$<$12\% for $>$2~\Mjup \ planets between 50--295~AU at the 95\% confidence level.

\subsubsection{MMT $L'$ and $M$-Band Survey of Nearby Sun-Like Stars}{\label{sec:youngaustral}}

\citet{Heinze:2010dm} carried out a deep $L'$- and $M$-band survey of 54 nearby FGK stars at the MMT with Clio.
The MMT adaptive optics system uses a deformable secondary mirror which reduces the thermal background by
minimizing the number of optical elements along the light path.  Observations were carried out between 2006--2007 with angular differential
imaging.  The image processing pipeline is described in \citet{Heinze:2008fg} and \citet{Heinze:2010dm}.
The target ages are generally older ($\sim$0.1--2~Gyr) but the long wavelengths of the observations and proximity of the sample ($\lesssim$25~pc) 
enabled sensitivity to planetary masses for most of the targets.
One new low-mass stellar companion was discovered and the binary brown dwarf HD~130948~BC was 
recovered in the survey.
The statistical results are detailed in \citet{Heinze:2010ko}; they find that no more than 50\% of Sun-like stars
host $\ge$5~\Mjup \ planets between 30--94~AU and no more than 15\% host $\ge$10~\Mjup \ planets between 22--100~AU at the 90\% confidence level.

\subsubsection{NaCo Survey of Young Nearby Austral Stars}{\label{sec:youngaustral}}

This program utilized NaCo at the VLT between 2002--2007 to target 88 young GKM
stars within 100~pc (\citealt{Chauvin:2010hm}).
17 new close multiple systems were uncovered and deep imaging was obtained for 65 single
young stars.
Observations were taken with a Lyot coronagraph in $H$ and $K_S$ bands and
PSF subtraction was performed with azimuthally-averaged subtraction and high-pass filtering.

The most important discovery from this survey was 2M1207--3932~b, 
a remarkable 5~$\pm$~2~\Mjup \ companion to a 25~\Mjup \ brown dwarf in the 10~Myr 
TWA moving group (\citealt{Chauvin:2004cy}; \citealt{Chauvin:2005gg}) enabled with infrared wavefront sensing.  
The unusually red colors and spectral shape of 2M1207--3932~b (\citealt{Patience:2010hf}) have made it the prototype of 
young dusty L dwarfs, now understood as a cloudy extension of the 
L dwarf sequence to low temperatures (\citealt{Barman:2011dq}; \citealt{Marley:2012fo}).
This system is also unusual from the perspective of brown dwarf demographics; the mass ratio of
$\sim$0.2 and separation of $\sim$41~AU make it an outlier compared to brown dwarf 
mass ratio and separation distributions in the field (\citealt{Burgasser:2007uf}).
Two other substellar companions were discovered in this survey:
GSC~08047-00232~B (\citealt{Chauvin:2005it}), also independently found by \citet{Neuhauser:2003tb}, 
and AB~Pic~B (\citealt{Chauvin:2005dh}),
which resides near the deuterium-burning limit.

\subsubsection{NaCo Survey of Young Nearby Dusty Stars}{\label{sec:youngdusty}}

This VLT/NaCo survey targeted 59 young nearby AFGK stars with ages $\lesssim$200~Myr and
distances within 65~pc (\citealt{Rameau:2013it}).  Most of the sample are members of young moving groups and the majority (76\%) were chosen to 
have mid-infrared excesses, preferentially selected for having debris disks.
Observations were carried out in $L'$-band between 2009--2012 using angular differential imaging.
Four targets in the sample had known substellar companions (HR~7329, AB~Pic, HR~8799, and $\beta$~Pic).
No new substellar companions were discovered but eight new visual binaries were resolved.
A statistical analysis of AF stars between 5--320~AU and 3--14~\Mjup \ implies a
giant planet occurrence rate of 7.4$^{+3.6}_{-2.4}$\% (68\% confidence level).

\subsubsection{SEEDS: Strategic Exploration of Exoplanets and Disks with Subaru}{\label{sec:seeds}}

The SEEDS survey (PI: M. Tamura) was a 125-night program on the 8.2-m Subaru Telescope targeting about 500 stars to search for 
giant planets and spatially resolve circumstellar disks (\citealt{Tamura:2009ip}).
\citet{Tamura:2016jg} provide an overview of the observing strategy, target samples, and main results.
Observations were carried out 
with the HiCIAO camera behind Subaru's AO188 adaptive optics system
over five years beginning in 2009.
The sample contained a mixture of young stars in star-forming regions, moving groups, and open clusters;
nearby stars and white dwarfs; and stars with protoplanetary disks and debris disks.
Most of the observations were taken in $H$-band in angular differential imaging mode as well as polarimetric 
differential imaging for young disk-bearing stars.
The ADI reduction pipeline is described in \citet{Brandt:2013in}.

Three new substellar companions were found in SEEDS: GJ~758~B (\citealt{Thalmann:2009ca}), 
$\kappa$~And~B (\citealt{Carson:2013fw}), and GJ~504~b (\citealt{Kuzuhara:2013jz}).
The masses of GJ~504~b and $\kappa$~And~B may
fall in the planetary regime depending on the ages and metallicities of the system, which are still under debate.
Two brown dwarf companions found in the Pleiades (HD~23514~B and HII~1348~B; \citealt{Yamamoto:2013gu})
had also independently been discovered by other groups.
SEEDS resolved a remarkable number of protoplanetary and transition disks in polarized light--- over two dozen in total ---
revealing previously unknown gaps, rings, and spiral structures down to 0$\farcs$1 with exceptional clarity 
(e.g., \citealt{Thalmann:2010hz}; \citealt{Hashimoto:2011bt}; \citealt{Mayama:2012ez}; \citealt{Muto:2012is}).

The statistical results for debris disks are presented in \citet{Janson:2013cjb}.
At the 95\% confidence level, they find that $<$15--30\% of stars host $>$10~\Mjup \ planets at the gap edge.
\citet{Brandt:2014cw} inferred a frequency of 1.0--3.1\% at the 68\% confidence level for 5--70~\Mjup \ 
companions between 10--100~AU by
combining results from the SEEDS moving group sample (\citealt{Brandt:2014hc}),
the SEEDS disk sample (\citealt{Janson:2013cjb}), the SEEDS Pleiades sample 
(\citealt{Yamamoto:2013gu}), GDPS (\citealt{Lafreniere:2007cv}), and
the NICI Campaign moving group sample (\citealt{Biller:2013fu}).

\subsubsection{Gemini NICI Planet-Finding Campaign}{\label{sec:nici}}
The Gemini NICI Planet-Finding Campaign (PI: M. Liu) was a 500-hour survey targeting about 230 young stars
of all spectral classes with deep imaging using the Near-Infrared Coronagraphic Imager on the
Gemini-South 8.1-m telescope (\citealt{Liu:2010hn}).
NICI is an imaging instrument encompassing an adaptive optics system, tapered and partly-translucent Lyot coronagraph,
and dual-channel camera (\citealt{Chun:2008et}).
Campaign observations spanned 2008--2012 and were carried out in two modes: single-channel 
$H$-band with angular differential imaging, and simultaneous dual-channel ($CH_4S$ at 1.578~$\mu$m and $CH_4L$ at 1.652~$\mu$m)
angular and spectral differential imaging to maximize sensitivity to methane-dominated planets.
The observing strategy and reduction pipeline are detailed in \citet{Biller:2008kk} and \citet{Wahhaj:2013fq},
and NICI astrometric calibration is discussed in \citet{Hayward:2014dk}.

One previously-known brown dwarf companion was resolved into a close binary, HIP~79797~Bab (\citealt{Nielsen:2013jy}), and
three new substellar companions were found: PZ~Tel~B, a highly eccentric brown dwarf companion in the $\beta$~Pic
moving group (\citealt{Biller:2010ku}); CD-35~2722~B, a young mid-L dwarf in the AB~Dor moving group (\citealt{Wahhaj:2011by}); 
and HD~1160~B, a substellar companion orbiting a young massive star (\citealt{Nielsen:2012jk}).
No new planets were discovered but $\beta$~Pic~b was recovered during the survey and its orbit was shown to be
misaligned with the inner and outer disks (\citealt{Nielsen:2014js}; \citealt{Males:2014jl}).
Two debris disks surrounding HR~4796~A and HD~141569 were also resolved with unprecedented detail (\citealt{Wahhaj:2014ur}; \citealt{Biller:2015bu}).

The statistical results are organized in several studies.
From a sample of 80 members of young moving groups, \citet{Biller:2013fu} measured the frequency
of 1--20~\Mjup \ planets between 10--150~AU to be $<$6--18\%  at the 95.4\% confidence level, depending on which hot-start evolutionary models
are adopted.  The high-mass sample of 70 B and A-type stars was described in \citet{Nielsen:2013jy}; they found that 
the frequency of $>$4~\Mjup \  planets between 59--460~AU is $<$20\% at 95\% confidence.
\citet{Wahhaj:2013iq} found that $<$13\% of debris disk stars have $\ge$5~\Mjup \ planets beyond 80~AU at 95\% confidence
from observations of 57 targets.

\subsubsection{IDPS: International Deep Planet Search}{\label{sec:idps}}

IDPS is an expansive imaging survey carried out at the VLT with NaCo, Keck with NIRC2, Gemini-South with NICI, and Gemini-North with NIRI targeting 
$\approx$300 young A--M stars (PI: C. Marois).  
This 14-year survey was mostly carried out in $K$ band, though much of the survey comprised a 
mix of broad- and narrow-band near-infrared filters.
Target ages were mostly $\lesssim$300~Myr and encompassed distances from $\sim$10--80~pc (Galicher et al. 2016, submitted).

The main result from this survey was the discovery of the HR~8799 planets (\citealt{Marois:2008ei}; \citealt{Marois:2010gpa}).
Altogether over 1000 unique point sources were found, most of which were meticulously shown to be 
background stars from multi-epoch astrometry (Galicher et al. 2016, submitted).
The preliminary analysis of a subset of high-mass A and F stars spanning $\approx$1.5--3.0~\Msun \
was presented in \citet{Vigan:2012jm}.
39 new observations in $H$, $K$, and $CH4_S$ filters were carried out in angular differential
imaging mode between 2007--2012 and were combined with three high-mass targets from the literature.
Stellar ages span 8--400~Myr with distances out to 90~pc and comprise a mix
of young moving group members, young field stars, and debris disk hosts.
The subsample of 42 massive stars includes three hosts of substellar companions:
HR~8799, $\beta$~Pic, and HR~7329, a $\beta$~Pic moving group member with a wide 
brown dwarf companion (\citealt{Lowrance:2000ic}).
Including the detections of planets around HR~8799 and $\beta$~Pic, \citet{Vigan:2012jm} measure the occurrence
rate of 3--14~\Mjup \ planets between 5--320~AU to be 8.7$^{+10.1}_{-2.8}$\% \ at 68\% confidence.

The complete statistical analysis for the entire sample is presented in Galicher et al. (2016, submitted).
They merge their own results for 292 stars with the GDPS and NaCo-LP surveys,
totaling a combined sample of 356 targets.  From this they infer an occurrence rate
of 1.05$^{+2.80}_{-0.70}$\% (95\% confidence interval) for 0.5--14~\Mjup \ planets
between 20--300~AU.  They do not find evidence that this frequency depends
on stellar host mass.  In addition, 16 of the 59 binaries resolved in IDPS are new.

\subsubsection{PALMS: Planets Around Low-Mass Stars}{\label{sec:palms}}

The PALMS survey (PI: B. Bowler) is a deep imaging search for planets and brown dwarfs orbiting low-mass stars
(0.1--0.6~\Msun) carried out at Keck Observatory with NIRC2 and Subaru Telescope with HiCIAO.  
Deep coronagraphic observations were acquired for 78 single young M dwarfs in $H$- and $Ks$-bands between 2010--2013 using 
angular differential imaging.  An additional 27 stars were found to be close binaries.
Targets largely originate from \citet{Shkolnik:2009dx}, \citet{Shkolnik:2012cs}, and an additional $GALEX$-selected sample
(E. Shkolnik et al., in preparation).  Most of these 
lie within 40~pc and have ages between 20--620~Myr; about one third of the sample are members
of young moving groups.
The observations and PSF subtraction pipeline are described in \citet{Bowler:2015ja}.

Four substellar companions were found in this program: 1RXS~J235133.3+312720~B (\citealt{Bowler:2012cs}),
GJ~3629~B (\citealt{Bowler:2012dc}), 1RXS~J034231.8+121622~B (\citealt{Bowler:2015ja}), 
and 2MASS~J15594729+4403595~B (\citealt{Bowler:2015ja}).
1RXS~J235133.3+312720~B is a particularly useful benchmark 
brown dwarf because it orbits a member of a young moving group (AB Dor) 
and therefore has a well-constrained age ($\approx$120~Myr).

The statistical results from the survey are presented in \citet{Bowler:2015ja}.
No planets were found, implying an occurrence rate of $<$10.3\% for 1--13~\Mjup \ 
planets between 10--100~AU at the 95\% confidence level assuming hot-start models and $<$16.0\% assuming
cold-start models.  For the most massive planets between 5--13~\Mjup, the upper limits
are $<$6.0\% and $<$9.9\% for hot- and cold-start cooling models.

The second, parallel phase of the PALMS survey is an ongoing program 
targeting a larger sample of $\sim$400 young M dwarfs primarily at Keck 
with shallower contrasts (Bowler et al., in prep.).  Initial discoveries include 
two substellar companions: 2MASS~J01225093--2439505~B, an L-type member of AB~Dor 
at the planet/brown dwarf boundary with an unusually red spectrum (\citealt{Bowler:2013ek}; \citealt{Hinkley:2015gk}),
and 2MASS~J02155892--0929121~C (\citealt{Bowler:2015ch}), a brown dwarf in a close quadruple
system which probably belongs to the Tuc-Hor moving group.

\subsubsection{NaCo-LP: VLT Large Program to Probe the Occurrence of Exoplanets and Brown Dwarfs at Wide Orbits}{\label{sec:vltlp}}

The NaCo-LP survey was a Large Program at the VLT focused on 86 young,
bright, primarily FGK stars (PI: J.-L. Beuzit).  $H$-band observations were carried out with NaCo in ADI mode
between 2009--2013 (\citealt{Chauvin:2015jy}).  The target sample is described in detail 
in \citet{Desidera:2015fl}; stars were chosen to be single, have ages $\lesssim$200~Myr, and 
lie within 100~pc.  Many of these stars were identified as new 
members of young moving groups.

Although no new substellar companions were discovered, 
an intriguing white dwarf was found orbiting
HD~8049, an ostensibly young K2 star that may instead be
much older due to mass exchange with its now evolved companion (\citealt{Zurlo:2013kb}).
New observations of the spatially resolved debris disk around 
HD~61005 (``the Moth'') were presented by \citet{Buenzli:2010ga},
and 11 new close binaries were resolved during this program (\citealt{Chauvin:2015jy}).

The statistical analysis of the sample of single stars was performed in \citet{Chauvin:2015jy}.
Based on a subsample of 51 young FGK stars, they found that $<$15\% of Sun-like stars host
planets with masses $>$5~\Mjup \ between 100--500~AU and
$<$10\% host $>$10~\Mjup \ planets between 50--500~AU at the 95\% confidence level.
\citet{Reggiani:2016dn} use these NaCo-LP null results together with additional deep archival 
observations to study the companion mass function as it relates to binary star formation and planet formation.
From their full sample of 199 Sun-like stars, they find that the results from direct imaging
are consistent with the superposition of the planet mass function determined from 
radial velocity surveys and the stellar companion 
mass ratio distribution down to 5~\Mjup, suggesting that many planetary-mass companions 
uncovered with direct imaging may originate from the tail of the brown dwarf mass distribution instead of being the most
massive representatives of the giant planet population.

\subsection{Other First Generation Surveys}{\label{sec:otherfirst}}

Several smaller, more focused surveys have also been carried out with angular differential imaging: \citealt{Apai:2008gk} 
targeted 8 debris disk hosts with NaCo using simultaneous differential imaging;
\citet{Ehrenreich:2010dc} imaged 38 high-mass stars primarily with NaCo;
\citet{Janson:2011hu} observed 15 B and A stars with NIRI;
\citet{Delorme:2012bq} presented observations of 16 young
M dwarfs in $L'$-band with NaCo;
\citet{Maire:2014il} targeted 16 young AFGK stars with NaCo's four-quadrant phase mask, simultaneous
differential imaging, and angular differential imaging;
\citet{Meshkat:2015dh} used NaCo's Apodizing Phase Plate coronagraph in $L'$ band to image
six young debris disk hosts with gaps as part of the Holey Debris Disk survey;
and \citet{Meshkat:2015iz} also used the APP at NaCo to image a sample of 13~A- and F-type main sequence
stars in search of planets, uncovering a probable brown dwarf around HD~984 (\citealt{Meshkat:2015hd}).
The Lyot Project is another survey with important contributions for its early use of coronagraphy behind an
extreme adaptive optics system (\citealt{Oppenheimer:2004go}).  This survey was carried out at the 3.6-m AEOS telescope equipped with a
941-actuator deformable mirror and targeted 86 nearby bright stars using angular differential imaging (\citealt{Leconte:2010ed}).

\subsection{The Second Generation: Extreme Adaptive Optics, Exceptional Strehl Ratios, and Optimized Integral Field Units}
 
The transition to second-generation planet-finding instruments began over the past few years.  
This new era is characterized by regular implementation of high-order (``extreme'') adaptive optics systems with thousands of
actuators and exceptionally low residual wavefront errors;
pyramid wavefront sensors providing better sensitivity and higher precision wavefront correction;
Strehl ratios approaching (and often exceeding) 90\% at near-infrared wavelengths;
high-contrast integral field units designed for on-axis observations enabling speckle subtraction and low-resolution spectroscopy;
sensitivity to smaller inner working angles than first-generation instruments; and advanced coronagraphy.

 \subsubsection{Project 1640}{\label{sec:leech}}

Project 1640 is a large ongoing survey (PI: R. Oppenheimer) and 
high-contrast imaging instrument with the same name located behind the PALM-3000 second-generation adaptive
optics system at the Palomar Observatory 200-inch (5.1-meter) Hale Telescope.
The instrument itself contains an apodized-pupil Lyot coronagraph and integral field unit that samples 32 spectral 
channels across $Y$, $J$, and $H$ bands (\citealt{Hinkley:2011kh}), producing a low-resolution 
spectrum to broadly characterize the physical properties of faint companions (\citealt{Roberts:2012et}; \citealt{Rice:2015kp}).
The survey consists of two phases, the first (now concluded) with the original 
Palomar Adaptive Optics System (PALM-241; \citealt{Troy:2000wp}) and a second ongoing three-year program 
focusing on nearby massive stars with the upgraded PALM-3000 adaptive optics system (\citealt{Dekany:2013hv}).
The data reduction pipeline is described in \citet{Zimmerman:2011iq} and a detailed treatment of
speckle subtraction, precision astrometry, and robust spectrophotometry can be found in
\citet{Crepp:2011kc}, \citet{Pueyo:2012ft}, \citet{Oppenheimer:2013gy}, \citet{Fergus:2014hv}, and \citet{Pueyo:2015cx}.

Results from the Project 1640 survey include several discoveries of faint stellar companions to massive
A stars (\citealt{Zimmerman:2010dn}; \citealt{Hinkley:2010id}) and follow-up astrometric and spectral characterization of
known substellar companions (\citealt{Crepp:2015gt}; \citealt{Hinkley:2013ko}).
In addition, \citet{Oppenheimer:2013gy} and \citet{Pueyo:2015cx} presented detailed spectroscopic and astrometric analysis of the 
 HR~8799 planets and found intriguing evidence for mutually dissimilar spectral properties and 
 signs of non-coplanar orbits.

\subsubsection{LEECH: LBTI Exozodi Exoplanet Common Hunt}{\label{sec:leech}}

LEECH is an ongoing $\sim$70-night high-contrast imaging program (PI: A. Skemer) at the twin 8.4-m Large Binocular Telescope.
Survey observations began in 2013 and are carried out in angular differential imaging mode in $L'$-band with LMIRcam
utilizing deformable secondary mirrors
to maximize sensitivity at mid-IR wavelengths by
limiting thermal emissivity from warm optics (\citealt{Skemer:2014ch}).
The target sample focuses on intermediate-age stars $<$1~Gyr including members of the $\sim$500~Myr Ursa Majoris moving group,
massive BA stars, and nearby young FGK stars.

In addition to searching for new companions, this survey is also characterizing known 
planets using the unique mid-IR instrumentation, sensitivity, 
and filter suite at the LBT.    \citet{Maire:2015dt} refined the orbits of the HR~8799 planets and found
them to be consistent with 8:4:2:1 mean motion resonances.
\citet{Skemer:2016ha} observed GJ~504~b in three narrow-band filters spanning 3.7--4.0 $\mu$m.  Model fits
indicate an exceptionally low effective temperature of $\approx$540~K and enhanced metallicity,
possibly pointing to an origin through core accretion.
Additionally, \citet{Schlieder:2016ih} presented LEECH observations and 
dynamical mass measurements the Ursa Majoris binary NO UMa.
Recently the integral field unit Arizona Lenslets for Exoplanet Spectroscopy (ALES; \citealt{Skemer:2015is})
was installed inside LMIRcam and will enable integral-field spectroscopy of planets between 3--5~$\mu$m
for the first time.

\subsubsection{GPIES: Gemini Planet Imager Exoplanet Survey}{\label{sec:gpies}}

GPIES is an ongoing 890-hour, 600-star survey to image extrasolar giant planets and debris disks
with the Gemini Planet Imager
at Gemini-South (PI: B. Macintosh).  GPI is expressly built to image planets at small
inner working angles;
its high-order adaptive optics system
incorporates an apodized pupil Lyot coronagraph, integral field spectrograph, 
imaging polarimeter, and (imminent) non-redundant masking capabilities (\citealt{Macintosh:2014js}).
Survey observations targeting young nearby stars began in 2014 and will span three years.  

\citet{Macintosh:2015ewa} presented the discovery of 51 Eri b, the first exoplanet found in GPIES
and the lowest-mass planet imaged in thermal emission to date. 
This remarkable young, methanated T dwarf has a contrast of 14.5~mag in $H$-band at a separation of 0$\farcs$45,
which translates into a mass of only 2~\Mjup \ at 13~AU assuming hot-start cooling models.
It is also the only imaged planet consistent with the most pessimistic cold-start evolutionary models, in which
case its mass may be as high as 12~\Mjup.  \citet{DeRosa:2015jla} obtained follow-up observations with GPI
and showed that 51~Eri~b shares a 
common proper motion with its host and exhibits slight (but significant) orbital motion.
Other initial results from this survey include astrometry and a refined orbit for $\beta$~Pic~b (\citealt{Macintosh:2014js}),
as well as resolved imaging of the debris disks around HD~106906 (\citealt{Kalas:2015en}) and 
HD~131835 (\citealt{Hung:2015dpa}).

\subsubsection{SPHERE GTO Survey}{\label{sec:sphere}}

SPHERE (Spectro-Polarimetric High-contrast Exoplanet Research) 
is an extreme adaptive optics system (SAXO) and versatile instrument for high-contrast imaging and spectroscopy 
at the VLT with a broad range of capabilities (\citealt{Beuzit:2008gt}).  IRDIS (\citealt{Dohlen:2008eu})
offers classical and dual-band imaging (\citealt{Vigan:2010cs}), dual-polarization imaging (\citealt{Langlois:2014dt}), 
and long-slit spectroscopy (\citealt{Vigan:2008he}); IFS provides low-resolution ($R$$\sim$30--50) 
integral field spectroscopy spanning 0.95--1.65~$\mu$m (\citealt{Claudi:2008dj}); and ZIMPOL enables diffraction-limited imaging
and polarimetry in the optical (\citealt{Thalmann:2008gi}).
As part of a guaranteed time observing program, the SPHERE GTO team is carrying out a large ongoing
survey (PI: J.-L. Beuzit) with a range of science goals.  About 200 nights of this time are devoted to a deep near-infrared imaging survey 
(SHINE: SpHere Infrared survEy) to search for 
and characterize exoplanets around 400--600 stars, while the remaining $\sim$60 nights will be used
for a broad range of science including spatially-resolved observations of circumstellar disks and optical imaging of planets.

Initial results from the SHINE survey include observations of the brown dwarf GJ~758~B (\citealt{Vigan:2016gq}),
PZ Tel B and HD~1160 B (\citealt{Maire:2016go}), HD~106906 (\citealt{Lagrange:2016bh}), 
and the HR~8799 planets (\citealt{Zurlo:2016hl}; \citealt{Bonnefoy:2016gx}).
Other results have focused on resolved imaging of 
the debris disk surrounding HD~61005, which may be
a product of a recent planetesimal collision (\citealt{Olofsson:2016ts}),
and HD~135344~B, host of a transition disk with striking spiral arm structure (\citealt{Stolker:2016ul}).
\citet{Boccaletti:2015ji} uncovered intriguing and temporally evolving features in AU~Mic's debris disk.
In addition, \citet{Garufi:2016vt} presented deep IRDIS near-infrared images and visible ZIMPOL polarimetric 
observations of HD~100546 revealing a complex disk environment with considerable structure
and resolved $K$-band emission at the location of the candidate protoplanet HD~100546~b.

\subsubsection{Other Second Generation Instruments and Surveys}{\label{sec:sphere}}

A number of other novel instruments and forthcoming surveys bear highlighting.
MagAO (PI: L. Close) at the Magellan 6.5-m Clay telescope is a versatile adaptive optics system consisting of
a 585-actuator adaptive secondary mirror, pyramid wavefront sensor, and two science
cameras offering simultaneous diffraction-limited imaging spanning the visible (0.6--1.05~$\mu$m) with VisAO and near-infrared (1--5.3~$\mu$m)
with Clio2 (\citealt{Close:2012ed}; \citealt{Morzinski:2014ep}; \citealt{Morzinski:2015gb}).  
Strehl ratios of $\sim$20--30\% in the optical are opening up new science fronts including deep
red-optical observations of exoplanets (\citealt{Males:2014jl}; \citealt{Wu:2015cz}), characterization of accreting protoplanets in H$\alpha$ (\citealt{Sallum:2015ej}),
and high spatial resolution imaging down to $\sim$20~mas (\citealt{Close:2013bu}).
Vector apodizing phase plate coronagraphs were recently installed in MagAO and 
other upgrades such as an optical integral field unit are possible in the future.

Subaru Coronagraphic Extreme Adaptive Optics (SCExAO; PI: O. Guyon) is being built for the Subaru telescope and is
the newest extreme adaptive optics system on a large telescope.  A detailed description of all facets of this instrument
is described in \citet{Jovanovic:2015ja}.
In short, a pyramid wavefront sensor is coupled 
with a 2000-element deformable mirror to produce Strehl ratios in excess of 90\%. 
The instrument is particularly flexible, allowing for a variety of setups and instrument subcomponents including
speckle nulling to suppress static and slowly changing speckles (\citealt{Martinache:2014jj}),
a near-infrared science camera (currently HiCIAO), sub-diffraction-limited interferometric 
science in the visible with VAMPIRES (\citealt{Norris:2015jw}) and FIRST (\citealt{Huby:2012bp}),
high-contrast integral field spectroscopy (\citealt{Brandt:2014jv}),
and coronagraphy with phase-induced amplitude apodization  (\citealt{Guyon:2003gw})
and vector vortex coronagraphs (\citealt{Mawet:2010el}).

\begin{figure*}
  \vskip -.8 in
  \hskip -.1 in
  \resizebox{7.8in}{!}{\includegraphics{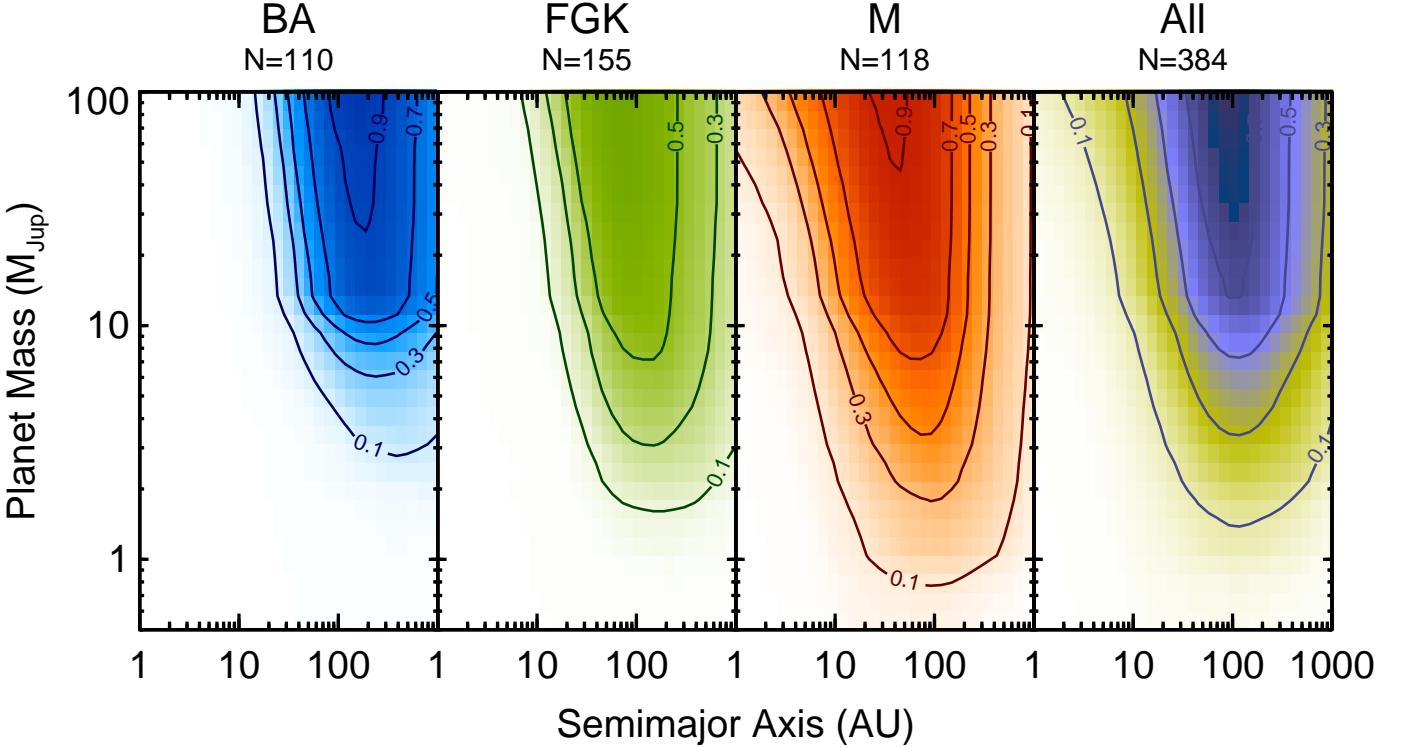}}
  \vskip -0.7 in
  \caption{Mean sensitivity maps from a meta-analysis of 384 unique stars with published high-contrast imaging observations.  M dwarfs provide the highest sensitivities to lower planet masses in the contrast-limited regime.  Altogether, current surveys probe the lowest masses at separations of $\sim$30--300~AU.  Contours denote 10\%, 30\%, 50\%, 70\%, and 90\% sensitivity limits.  \label{fig:meansensitivity} } 
\end{figure*}

\begin{figure}
  \vskip -.2 in
  \hskip -.6 in
  \resizebox{4.1in}{!}{\includegraphics{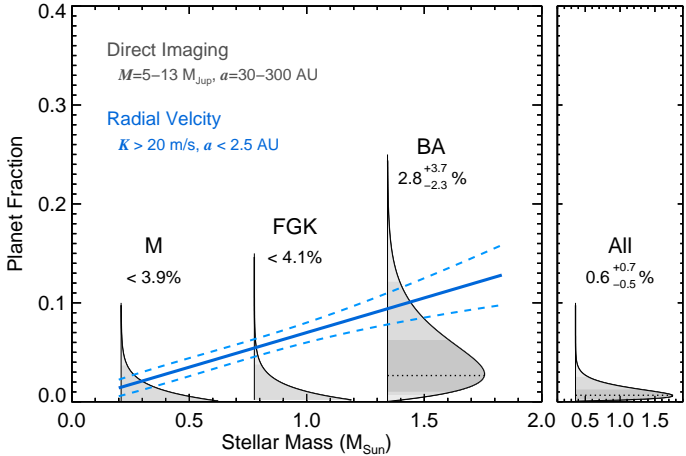}}
  \vskip -0.2 in
  \caption{Probability distributions for the occurrence rate giant planets  
  from a meta-analysis of direct imaging surveys in the literature.
   2.8$^{+3.7}_{-2.3}$\% of BA stars, $<$4.1\% of FGK stars, and $<$3.9\% of M dwarfs harbor giant planets
   between  5--13~\Mjup \  and 30--300~AU.  The correlation between stellar host mass and giant planet
   frequency at small separations ($<$2.5~AU) from \citet{Johnson:2010gu} is shown in blue.  Larger sample sizes are needed
   to discern any such correlation on wide orbits.  
   0.6$^{+0.7}_{-0.5}$\% of stars of any mass host giant planets over the same mass and separation range. \label{fig:plfrequency} } 
\end{figure}

\subsection{The Occurrence Rate of Giant Planets on Wide Orbits: Meta-Analysis of Imaging Surveys}{\label{sec:occurrencerate}}

The frequency and mass-period distribution of planets spanning various orbital distances, stellar host masses, and system ages 
provides valuable clues about the dominant processes shaping the 
formation and evolution of planetary systems.  
These measurements are best addressed with large samples and uniform statistical analyses.
\citet{Nielsen:2008kk} carried out the first such large-scale study based on  
adaptive optics imaging surveys from \citet{Biller:2007ht} and \citet{Masciadri:2005gl}.
From their sample of 60~unique stars they found an upper limit of 20\% for $>$4~\Mjup \ planets between 20--100~AU
at the 95\% confidence level.
This was expanded to 118 targets in \citet{Nielsen:2010jt} by including the GDPS survey of \citet{Lafreniere:2007cv},
resulting in the same upper limit and planet mass regime but for a broader range of separations of 8--911~AU at 68\% confidence. 
\citet{Vigan:2012jm} and \citet{Rameau:2013it} combined their own observations of high-mass stars with previous surveys
and measured occurrence rates of  8.7$^{+10.1}_{-2.8}$\% (for 3--14~\Mjup \ planets between 5--320~AU) and
16.1$^{+8.7}_{-5.3}$\% (for 1--13~\Mjup \ planets between 1--1000~AU), respectively.
\citet{Brandt:2014cw} incorporated the SEEDS, GDPS, and the NICI moving group surveys and found
a frequency of 1.0--3.1\% for 5--70~\Mjup \ companions between 10--100~AU.
Recently, Galicher et al. (2016, submitted) combined results from IDPS, GDPS, and the
NaCo Survey of Young Nearby Austral Stars and found an occurrence rate of 1.05$^{+2.80}_{-0.70}$\%
for 0.5--14~\Mjup \ companions between 20--300~AU based on a sample of 356 unique stars.  
Breaking this into stellar mass bins did not reveal
any signs of a trend with stellar host mass.

\begin{figure*}
  \vskip -1.2 in
  \hskip -.2 in
  \resizebox{7.3in}{!}{\includegraphics{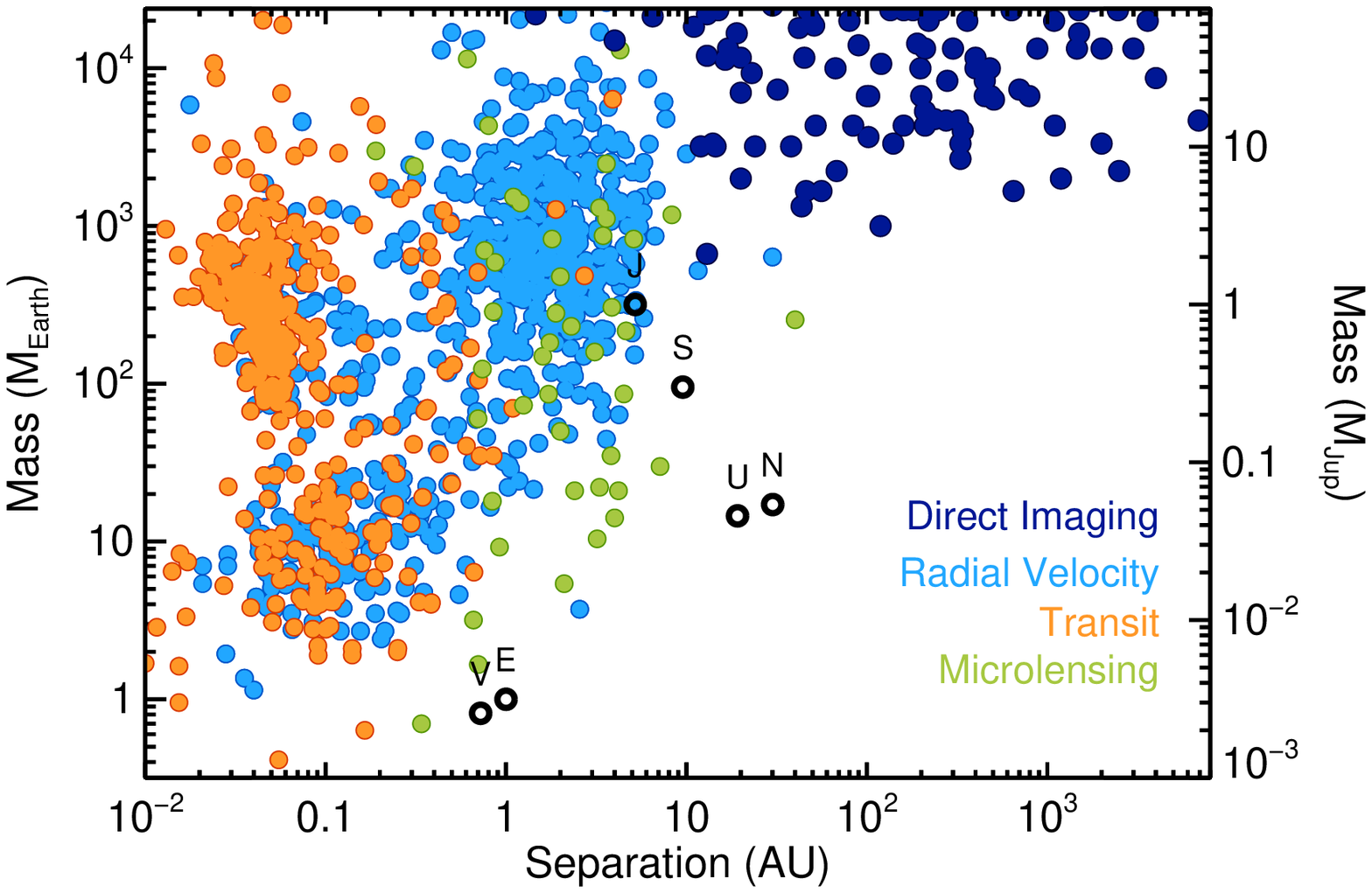}}
  \vskip -.6 in
  \caption{The demographics of exoplanets from direct imaging (dark blue), radial velocity (light blue), transit (orange), and microlensing (green) surveys.  Planets detected with radial velocities are minimum masses.  It remains unclear whether imaged planets and brown dwarfs represent distinct populations or whether they form a continuous distribution down to the fragmentation limit.  Directly imaged substellar companions are compiled from the literature, while planets found with other methods are from exoplanets.eu as of April 2016. \label{fig:mass_sma_exoplanets} } 
\end{figure*}

Here I reexamine the occurrence rate of giant planets with a meta-analysis of 
the largest and deepest high-contrast imaging surveys.
696 contrast curves are assembled from the literature from the programs 
outlined in Section~\ref{sec:firstgen}.  For stars with more than one observation,
the deeper contrast curve at 1$''$ is chosen.  Targets with stellar companions within 100 AU are removed from
the sample because binaries can both inhibit planet formation and dynamically disturb planetary orbits.  
Most candidate planets uncovered during these surveys are rejected as background stars
from second epoch observations, but some candidates are either not recovered or 
are newly revealed in follow-up data.  Because of finite telescope allocation,
some of these candidates remain untested for common proper motion.
These ambiguous candidates cannot  
be ignored in a statistical analysis because one (or more) could be indeed be bound.
In these cases, contrast curves are individually truncated 
one standard deviation above the brightest candidate.
Ages are taken from the literature except for members of young moving groups, for which 
the most recent (and systematically older) ages of young
moving groups from \citet{Bell:2015gw} are adopted.
Most ages in the sample are less than 300~Myr and within 100~pc.
Altogether this leaves 384 unique stars spanning B2--M6 spectral types: 
76 from \citet{Bowler:2015ja}, 72 from \citet{Biller:2013fu}, 61 from \citet{Nielsen:2013jy}, 54 from \cite{Lafreniere:2007cv},
45 from \citet{Brandt:2014hc}, 30 from \citet{Janson:2013cjb}, 25 from \citet{Vigan:2012jm},
14 from \citet{Wahhaj:2013iq}, and 7 from \citet{Janson:2011hu}.

Sensitivity maps and planet occurrence rates are derived following \citet{Bowler:2015ja}.
For a given planet mass and semimajor axis, a population of artificial planets on random circular orbits 
are generated in a Monte Carlo fashion and converted into apparent magnitudes and 
separations using Cond hot-start evolutionary models
from \citet{Baraffe:2003bj}, the age of the host star, and the distance to the system, including 
uncertainties in age and distance.  
These are compared with the measured contrast curve to infer the fractional sensitivity at each grid point 
spanning 30 logarithmically-uniform bins in mass and separation between 1--1000~AU and 0.5--100~\Mjup.
When available, fractional field of view coverage is taken into account.
Contrasts measured in $CH_4S$ filters are converted to $H$-band using an empirical color-spectral type
relationship based on synthetic colors of ultracool dwarfs from the SpeX Prism Library (\citealt{Burgasser:2014tr}) as well as 
the spectral type-effective temperature sequence
from \citet{Golimowski:2004en}\footnote{Synthesized colors of ultracool 
dwarfs using the Keck/NIRC2 
$CH_4S$ and $H_\mathrm{MKO}$ filter profiles yields the following relation: 
$CH_4S$--$H_\mathrm{MKO}$ = $\sum_{i=0}^{4}$$c_i$SpT$^i$, where $c_0$=0.03913178, 
$c_1$=0.008678245, $c_2$=--0.001542768, $c_3$=0.0001033761, $c_4$=--2.902588$\times$10$^{-6}$,
and SpT is the numerical near-infrared spectral type (M0=1.0, L0=10.0, T0=20.0).
This relation is valid from M3--T8 and the rms of the fit is 0.025~mag.
 \citet{Golimowski:2004en} provide an empirical $T_\mathrm{eff}$(SpT) relationship, but the inverse SpT($T_\mathrm{eff}$) is necessary
for this filter conversion at a given mass and age.  Refitting the same data from Golimowski et al. yields the following:
SpT = $\sum_{i=0}^{4}$$c_i$$T_\mathrm{eff}$$^i$, where 
$c_0$=36.56779, $c_1$=--0.004666549, $c_2$=--9.872890$\times$10$^{-6}$, 
$c_3$=4.108142$\times$10$^{-09}$, $c_4$=-- 4.854263$\times$10$^{-13}$.
This is valid for 700~K $<$ $T_\mathrm{eff}$ $<$ 3900~K, the rms is 1.89~mag, and SpT
is the same numerical near-infrared spectral type as above.}.

The mean sensitivity maps for all 384 targets and separate bins containing BA stars (110 targets), 
FGK stars (155 targets), and M dwarfs (118 targets) are shown in Figure~\ref{fig:meansensitivity}.
In general, surveys of high-mass stars probe higher planet masses than deep imaging around M dwarfs
owing to differences in the host stars' intrinsic luminosities.  The most sensitive region for all stars is between
$\sim$30--300~AU, with less coverage at extremely wide separations because of limited fields of view
and at small separations in contrast-limited regimes.

The occurrence rate of giant planets for all targets and for each stellar mass bin are listed in Table~\ref{tab:gpfreq}, which assumes 
logarithmically-uniform distributions in 
mass and separation (see Section 6.5 of \citealt{Bowler:2015ja} for details).
The mode and 68.3\% minimum credible interval (also known as the highest posterior density interval) of the planet frequency probability distribution are reported.
Two massive stars in the sample host planets that were either discovered or successfully 
recovered in these surveys: $\beta$~Pic, with a
planet at 9~AU, and HR~8799, with planets spanning 15--70~AU.
HR~8799 is treated as a single detection.
The most precise occurrence rate measurement is between 5--13~\Mjup \ and 30--300~AU.
Over these ranges, the frequency of planets orbiting
BA, FGK, and M stars is 2.8$^{+3.7}_{-2.3}$\%, $<$4.1\%, and $<$3.9\%, respectively
(Figure~\ref{fig:plfrequency}).
Here upper limits are 95\% confidence intervals.
Although there are hints of a higher giant planet occurrence rate around massive stars 
analogous to the well-established correlation at small separations 
(\citealt{Johnson:2007et}; \citealt{Lovis:2007cy}; \citealt{Johnson:2010gu}; \citealt{Bowler:2010eo}), 
this trend is not yet statistically significant at wide orbital distances
and requires larger sample sizes in each stellar mass bin to unambiguously test this correlation.
Marginalizing over stellar host mass, the overall giant planet occurrence rate for the full sample of 384 stars
is 0.6$^{+0.7}_{-0.5}$\%, which happens to be comparable to the frequency of hot Jupiters around 
FGK stars in the field (1.2~$\pm$~0.4\%; \citealt{Wright:2012kma}) and in the $Kepler$ sample
(0.5~$\pm$~0.1\%; \citealt{Howard:2012di}).
 However, compared to the high occurrence rate of giant planets (0.3--10~\Mjup) with orbital periods out to 2000~days ($\sim$10\%; \citealt{Cumming:2008hg}),
massive gas giants are clearly quite rare at wide orbital distances.

\section{Brown Dwarfs, Giant Planets, and the Companion Mass Function}{\label{sec:dbl}

Direct imaging has shown that planetary-mass companions exist at unexpectedly wide separations
but the provenance of these objects remains elusive.
There is substantial evidence that the tail-end of the star formation process can produce objects 
extending from low-mass stars at the hydrogen burning limit ($\approx$75~\Mjup)
to brown dwarfs at the opacity limit for fragmentation 
($\approx$5--10~\Mjup), which corresponds to the minimum mass of a pressure-supported fragment 
during the collapse of a molecular cloud core 
(\citealt{Low:1976wt}; \citealt{Silk:1977il}; \citealt{Boss:2001vw}; \citealt{Bate:2002iq}; \citealt{Bate:2009br}).
Indeed, isolated objects with inferred masses below 10~\Mjup \ have been found in a range of contexts over the past decade:
in star-forming regions (\citealt{Lucas:2001ed}; \citealt{Luhman:2009cn}; \citealt{Scholz:2012kv}; \citealt{Muzic:2015kn}), 
among closer young stellar associations (\citealt{Liu:2013gya}; \citealt{Gagne:2015kf}; \citealt{Kellogg:2016fo}; \citealt{Schneider:2016iq}), 
and at much older ages as Y dwarfs in the field (\citealt{Cushing:2011dk}; \citealt{Kirkpatrick:2012ha}; \citealt{Beichman:2013dl}).
Similarly, several systems with \emph{companions} below $\approx$10~\Mjup \ are difficult to explain with any formation scenario other than
cloud fragmentation: 2M1207--3932~Ab is a $\sim$25~\Mjup \ brown dwarf with a $\sim$5~\Mjup \ companion at an 
orbital distance of 40~AU (\citealt{Chauvin:2004cy})
and 2M0441+2301 AabBab is a quadruple system comprising a low-mass star, two brown dwarfs, and a 
10~\Mjup \ object in a hierarchical and distinctly non-planetary configuration (\citealt{Todorov:2010cn}).  

From the radial velocity perspective, the distribution of gas giant minimum masses 
is generally well-fit with a decaying power law (\citealt{Butler:2006dd}; \citealt{Johnson:2009iz}; \citealt{Lopez:2012jp}) 
or exponential function (\citealt{Jenkins:2016um}) that tapers off beyond $\sim$10~\Mjup.
This is evident in Figure~\ref{fig:mass_sma_exoplanets}, although inhomogeneous radial velocity detection biases 
which exclude lower-mass planets at wide separations are not taken into account.  
The dominant formation channel for this population of close-in giant planets is thought to be core accretion plus gas capture,
in which growing cores reach a critical mass and undergo runaway gas accretion (e.g., \citealt{Helled:2013et}).

The totality of evidence indicates that 
the decreasing brown dwarf companion mass function almost certainly overlaps with the the rising giant planet mass function in the
5--20~\Mjup \ mass range.  
No strict mass cutoff can therefore unambiguously divide giant planets from brown dwarfs,
and many of the imaged companions below 13~\Mjup \  listed in Table~\ref{tab:planets} 
probably originate from the dwindling brown dwarf companion mass function.

Another approach to separate these populations is to consider 
formation channel: planets originate in disks while brown dwarfs form like stars from 
the gravitational collapse of molecular cloud cores.
However, not only are the relic signatures of formation difficult to discern for individual discoveries, but 
objects spanning the planetary up to the stellar mass regimes may also form in large Toomre-unstable circumstellar disks at
separations of tens to hundreds of AU (e.g., \citealt{Durisen:2007wg}; \citealt{Kratter:2016dn}).
Any binary narrative based on origin in a disk versus a cloud core is therefore also problematic.
Furthermore, both giant planets and brown dwarf companions may migrate, 
dynamically scatter, or undergo periodic Kozai-Lidov orbital oscillations if a third body is present, further mixing these
populations and complicating 
the interpretation of very low-mass companions uncovered with direct imaging.

The deuterium-burning limit at $\approx$13~\Mjup \ is generally acknowledged as a nebulous, imperfect, 
and ultimately artificial division
between brown dwarfs and giant planets.
Moreover, this boundary is not fixed and may 
depend on planet composition, core mass, and accretion history 
(\citealt{Spiegel:2011ip}; \citealt{Bodenheimer:2013ki}; \citealt{Mordasini:2013cr}).
Uncertainties in planet luminosities, 
evolutionary histories, metallicities, and ages can also produce large systematic errors in 
inferred planet masses (see Section~\ref{sec:masses}), rendering 
inconsequential any sharp boundary set by mass.
However, despite these shortcomings, this border lies in the planet/brown dwarf ``mass valley" and 
may still serve as a pragmatic (if flawed) 
qualitative division between two populations
formed \emph{predominantly}
with their host stars and \emph{predominantly} in protoplanetary disks.  

Observational tests of formation routes will eventually provide the necessary tools to understand the 
relationship between these populations.  
This can be carried out at an individual level with environmental clues such as coplanarity of multi-planet systems   
or orbital alignment within a debris disk; enhanced metallicities or abundance ratios relative to host stars (\citealt{Oberg:2011je});
or overall system orbital architecture.
Similarly, the statistical properties of brown dwarfs and giant planets can be used to identify dominant formation channels: 
the separation distribution of 
objects formed through cloud fragmentation should resemble that of binary stars; disk instability and core accretion 
may result in a bimodal period distribution for giant planets (\citealt{Boley:2009dk}); 
planet scattering to wide orbits should produce a rising
mass function at low planet masses as opposed to a truncated mass distribution at the fragmentation limit for 
cloud fragmentation and disk instability;  and the companion mass function and mass ratio distribution
are expected to smoothly extend from low-mass stars down to the fragmentation 
limit if a common formation channel in at play (\citealt{Brandt:2014cw}; \citealt{Reggiani:2016dn}). 
Testing these scenarios will require much larger sample sizes given the low occurrence rates 
uncovered in direct imaging surveys.
}

\section{Conclusions and Future Outlook}{\label{sec:intro}}

High-contrast imaging is still in its nascence.  Radial velocity, transit, and microlensing
surveys have unambiguously demonstrated that giant planets are much rarer than super-Earths
and rocky planets at separations $\lesssim$10~AU.  
In that light, the discovery of truly massive planets at tens, hundreds, and even thousands of AU
with direct imaging is fortuitous, even if the overall occurrence rate of this population is quite low.
Each detection technique has produced many micro paradigm shifts over the past
twenty years that disrupt and rearrange perceptions about the demographics and architectures
of planetary systems.  
Hot Jupiters, correlations with stellar mass and metallicity, 
the ubiquity of super-Earths, compact systems of small planets,
resonant configurations, orbital misalignments, 
the prevalence of habitable-zone Earth-sized planets, 
circumbinary planets, and featureless clouds and hazes 
are an incomplete inventory within just a few AU (e.g., \citealt{Winn:2015jt}).
The most important themes to emerge from direct imaging are that 
massive planets exist but are uncommon at wide separations ($>$10~AU), 
and at young ages the low-gravity atmospheres of giant planets do not resemble 
those of older, similar-temperature brown dwarfs.

There are many clear directions forward in this field.  Deeper contrasts and smaller
inner working angles will probe richer portions of planetary mass- and separation
distributions.  Thirty meter-class telescopes with extreme adaptive optics systems
will regularly probe sub-Jovian masses at separations down to 5~AU.
This next generation will uncover more planets and enable a complete mapping of the evolution of
giant planet atmospheres over time.
Other fertile avenues for high-contrast imaging include 
precise measurements of atmospheric composition (\citealt{Konopacky:2013jvc}; \citealt{Barman:2015dy}),
doppler imaging (\citealt{Crossfield:2014es}; \citealt{Crossfield:2014cy}),
photometric monitoring to map variability of rotationally-modulated features (e.g., \citealt{Apai:2016ky}),
synergy with other detection methods  
(e.g., \citealt{Lagrange:2013gh}; \citealt{Sozzetti:2013de}; \citealt{Montet:2014fa}; \citealt{Clanton:2016ft}),
advances in stellar age-dating at the individual and population levels,
merging high-contrast imaging with high-resolution spectroscopy (\citealt{Snellen:2014kz}; \citealt{Snellen:2015bta}),
surveying the companion mass function to sub-Jovian masses,
polarimetric observations of photospheric clouds  (e.g., \citealt{Marley:2013vq}; \citealt{JensenClem:2016kt}),
statistical correlations with stellar host properties,
probing the earliest stages of protoplanet assembly (\citealt{Kraus:2012gk}; \citealt{Sallum:2015ej}),
astrometric orbit monitoring and constraints on dynamical histories,
and robust dynamical mass measurements to test evolutionary models and 
probe initial conditions (e.g., \citealt{Dupuy:2009jq}; \citealt{Crepp:2012eg}).
High-contrast imaging has a promising future and will play an ever-growing role
in investigating the architecture, atmospheres, and origin of exoplanets.

\acknowledgments

It is a pleasure to thank the referee, Rebecca Oppenheimer, as well as 
Lynne Hillenbrand, Dimitri Mawet, Sasha Hinkley, and Trent Dupuy for their thoughtful comments and constructive feedback on this review.
Michael Liu, Arthur Vigan, Christian Marois, Motohide Tamura, Ga{\"e}l Chauvin, 
Andy Skemer, Adam Kraus, and Bruce Macintosh contributed helpful suggestions on past and ongoing imaging surveys.
Bruce Macintosh, Eric Nielsen, Andy Skemer, and Rapha{\"e}l Galicher kindly provided
images for Figure~\ref{fig:planetgallery}.  Trent Dupuy generously shared his compilation
of late-T and Y dwarfs used in Figure~\ref{fig:cmd}.
This research has made use of the Exoplanet Orbit Database,
the Exoplanet Data Explorer at exoplanets.org,
and the SpeX Prism Spectral Libraries maintained by Adam Burgasser.
 NASA's Astrophysics Data System Bibliographic Services together with the VizieR catalogue access tool and SIMBAD database 
operated at CDS, Strasbourg, France, were invaluable resources for this work.

\newpage


\clearpage

\newpage

\begin{deluxetable}{lcccccccccc}
\tabletypesize{\footnotesize}
\tablewidth{0pt}
\tablecolumns{11}
\tablecaption{Directly Imaged Planets and Planet Candidates with Inferred Masses $\lesssim$13~\Mjup \label{tab:planets}}
\tablehead{
        \colhead{Name}     &   \colhead{Mass\tablenotemark{a}}     &  \colhead{Luminosity}     &    \colhead{Age}              &    \colhead{Sep.}    &  \colhead{Sep.}    &   \colhead{NIR SpT}  & \colhead{Orbital}  & \colhead{Pri.}   & \colhead{Pri. Mass}  &  \colhead{References}                                              \\
        \colhead{}                &   \colhead{(\Mjup)}   & \colhead{(log~($L_\mathrm{Bol}$/$L_{\odot}$))}  &  \colhead{(Myr)}  &    \colhead{($''$)}               & \colhead{(AU)}               &                                       &  \colhead{Motion?}  &  \colhead{Mult.\tablenotemark{b}}  & \colhead{(\Msun)}     &  \colhead{} 
        }   
\startdata
\cutinhead{Close-in Planets ($<$100~AU)}
51 Eri b                                  &    2 $\pm$ 1  & --5.6 $\pm$ 0.2  &   23 $\pm$ 3   &  0.45  &  13   & T4.5--T6              &  Yes  &  S  &   1.75   &   1, 2, 3 \\
HD 95086 b                          &    5 $\pm$ 2  &  --4.96 $\pm$ 0.10  &  17 $\pm$ 4   & 0.6   &  56   & L/T:                   &  Yes   &  S  &   1.6     &  4--7  \\
HR~8799~b                          &  5 $\pm$ 1    &   --5.1 $\pm$ 0.1 & 40 $\pm$ 5   &  1.7    &  68   & $\sim$L/Tpec    & Yes  &  S  &  1.5  &  8--11  \\
LkCa 15 b\tablenotemark{c}  &   6 $\pm$ 4  &  $\cdots$           &   2 $\pm$ 1   &  0.08   &  20   &  $\cdots$             &  Yes  &  S  &  1.0  &  12--15      \\
HR~8799~c                          &   7 $\pm$ 2    &   --4.7 $\pm$ 0.1  & 40 $\pm$ 5  &  0.95   &  38   &  $\sim$L/Tpec   & Yes  & S   &  1.5  &  8--11  \\
HR~8799~d                          &   7 $\pm$ 2    &  --4.7 $\pm$ 0.2  &  40 $\pm$ 5  &  0.62   &  24   &  $\sim$L7pec    & Yes  &  S  &  1.5  &  8, 10, 11  \\
HR~8799~e                          &   7 $\pm$ 2    &  --4.7 $\pm$ 0.2  &  40 $\pm$ 5  &  0.38    & 14   & $\sim$L7pec     &  Yes  &  S  &  1.5  &  10, 11, 16  \\
$\beta$ Pic b                        &    12.7 $\pm$ 0.3  &  --3.78 $\pm$ 0.03 & 23 $\pm$ 3  &  0.4  &  9    & L1                 &  Yes  &  S  &   1.6 &  17--20 \\
\enddata
\end{deluxetable}

\clearpage

\begin{deluxetable}{lcccccccc}
\tabletypesize{\footnotesize}
\tablewidth{0pt}
\tablecolumns{9}
\tablecaption{Directly Imaged Stellar and Substellar Companions Inducing Shallow Radial Velocity Trends \label{tab:trends}}
\tablehead{
        \colhead{Name}     &   \colhead{Separation}     &  \colhead{Separation}     &  \colhead{Nature of}  &  \colhead{Comp. Mass}  &    \colhead{Companion}    & \colhead{Planet}  &   \colhead{d$v$/d$t$}  &  \\
        \colhead{}                &   \colhead{($''$)}                 & \colhead{(AU)}                  &  \colhead{Companion}   &  \colhead{(\Msun)}             &    \colhead{SpT}           &   \colhead{Host?}  &  \colhead{(m s$^{-1}$ yr$^{-1}$)} &  \colhead{References}   
        }   
\startdata
HD 19467 B    &   1.65       &   51   &  BD    &    0.050  &    T5.5         &  No    & --1.37 $\pm$ 0.09  & 1, 2 \\
HD 4747 B      &    0.60       &   11.3  &  BD  &   0.057  &   $\cdots$   &  No   & --30.7 $\pm$ 0.6\tablenotemark{a}   &  3  \\  
HR 7672 B         &   0.5        &   18    &   BD      &  0.066  &  L4.5     &   No    &  --24 $\pm$ 0.6\tablenotemark{a} &  4, 5 \\  
HIP 71898 B        &   3.0         &    30   &   LMS    &   0.074  &  L0        &  No   &  8.6 $\pm$ 0.4  &  6, 7, 8 \\ 
HD 5608 B         &   0.6         &   34     &  LMS     &  0.10    &  $\cdots$       &   Yes   &  --5.51 $\pm$ 0.45  &  9 \\  
HD 68017 B     &   0.59       &  13     &  LMS  &    0.15  &  [M5]        &   No    & 16.3 $\pm$ 0.9\tablenotemark{a}  &  10 \\  
HD 7449 B     &   0.54       &   18   &  LMS  &   0.2     &   [M4.5]        &  Yes  &   $\cdots$\tablenotemark{a}  &  11  \\ 
Gl 15 B              &   41          &    146   &   LMS   &   0.2      &  M3.5         &   Yes   &  --0.27 $\pm$ 0.09  &  12  \\
\enddata
\tablenotetext{a}{Radial velocities show both acceleration and curvature.}
\tablerefs{
(1) \citet{Crepp:2014ce}; 
(2) \citet{Crepp:2015gt}; 
(3) \citet{Crepp:2016ta}; 
(4) \citet{Liu:2002fx}; 
(5) \citet{Crepp:2012eg}; 
(6) \citet{Montet:2014fa}; 
(7) \citet{Golimowski:2004ba}
(8) \citet{Forveille:2004es};
(9) \citet{Ryu:2016vc};
(10) \citet{Crepp:2012bs}; 
(11) \citet{Rodigas:2016cf};
(12) \citet{Howard:2014bu};
(13) \citet{Schnupp:2010ts};
(14) \citet{Dupuy:2016dh};
(15) \citet{Bryan:2016cl};
(16) \citet{Howard:2010dia};
(17) \citet{Ginski:2016ea};
(18) \citet{Knutson:2014is};
(19)  \citet{Ngo:2015hn};
(20) \citet{Santos:2002ja};
(21)  \citet{Zucker:2003jd};
(22) \citet{Zucker:2004ka};
(23)  \citet{Mugrauer:2004daa};
(24) \citet{Neuhauser:2007hl};
(25) \citet{Crepp:2013io};
(26) \citet{Chauvin:2007bg};
(27) \citet{Chauvin:2011gf};
(28) \citet{Crepp:2013ij};
(29) \citet{Matthews:2014ee};
(30) \citet{Zurlo:2013kb};
(31) \citet{Queloz:2000tv};
(32) \citet{Els:2001hd};
(33) \citet{Mugrauer:2005kt};
(34) \citet{Chauvin:2006hm};
(35) \citet{Lagrange:2006ep}; 
(36) \citet{Farihi:2013ht};
(37) \citet{Irwin:1992bb}; 
(38) \citet{Torres:1999gc};
(39) \citet{Bond:2015do};
(40) \citet{Crepp:2012bs}.
}
\end{deluxetable}

\clearpage
\newpage

\begin{deluxetable}{lccccc}
\tabletypesize{\footnotesize}
\tablewidth{0pt}
\tablecolumns{6}
\tablecaption{The Frequency of 5--13~\Mjup \ Planets on Wide Orbits \label{tab:gpfreq}}
\tablehead{
        \colhead{Sample}   & \colhead{Number of}  &   \colhead{Occurrence Rate}     &  \colhead{Occurrence Rate}  &  \colhead{Occurrence Rate}  &  \colhead{Occurrence Rate}      \\
        \colhead{}                & \colhead{Stars}  &   \colhead{(10--100 AU)}    &   \colhead{(30--300 AU)}        & \colhead{(10--1000 AU)}    & \colhead{(100--1000 AU)}     
        }   
\startdata
BA & 110 &  7.7$^{+9.0}_{-6.0}$\% & 2.8$^{+3.7}_{-2.3}$\%  & 3.5$^{+4.7}_{-2.5}$\%  & $<$6.4\% \\ 
FGK & 155 &   $<$6.8\% &   $<$4.1\%  & $<$5.8\%  & $<$5.1\% \\
M & 118 &  $<$4.2\% &  $<$3.9\% & $<$5.4\%  & $<$7.3\% \\
All Stars & 384 & 0.8$^{+1.2}_{-0.5}$\% & 0.6$^{+0.7}_{-0.5}$\% & 0.8$^{+1.0}_{-0.6}$\% & $<$2.1\%  
\enddata
\tablecomments{Assumes circular orbits, logarithmically-flat planet mass-period distributions, and hot-start evolutionary models from \citet{Baraffe:2003bj}.  All binaries within 100~AU of the host stars have been removed.   Occurrence rates are 68\% credible intervals and upper limits are 95\% confidence values.}
\end{deluxetable}

\newpage

\clearpage

\appendix
\section{Astrometry of HR~8799 bcde and $\beta$~Pic~b}

\LongTables
\begin{deluxetable}{lccccccc}
\tabletypesize{\scriptsize}
\setlength{ \tabcolsep } {.1cm} 
\tablewidth{0pt}
\tablecolumns{8}
\tablecaption{Astrometry of HR 8799 bcde and $\beta$ Pic b Until 2016\label{tab:astrometry}}
\tablehead{
\colhead{UT Date}  &  \colhead{Epoch}  &  \colhead{Telescope/}  &  \colhead{$\rho$}  & \colhead{$\theta$}   & \colhead{$\Delta_\mathrm{R.A.}$} & \colhead{$\Delta_\mathrm{Dec}$}        & \colhead{Reference}  \\
\colhead{}  &  \colhead{}  &  \colhead{Instrument}  &  \colhead{($''$)}    & \colhead{($^{\circ}$)}      & \colhead{($''$)}                                & \colhead{($''$)}                          &  \colhead{}      
        }       
\startdata
\cutinhead{HR 8799 b}
1998 Oct 30  &  1998.828  &  $HST$/NICMOS &  1.721  $\pm$  0.012  &  55.1 $\pm$ 0.4    &  1.411  $\pm$ 0.009  &  0.986  $\pm$ 0.009  &  \citet{Lafreniere:2009dc}  \\
1998 Oct 30\tablenotemark{a}  &  1998.828  &  $HST$/NICMOS &  1.738  $\pm$  0.018  &  54.7 $\pm$ 0.4    &  [1.418  $\pm$ 0.016]  &  [1.00  $\pm$ 0.014]  &  \citet{Soummer:2011fk}  \\
2002 Jul 19  &  2002.545  &  Subaru/CIAO  &  [1.74  $\pm$  0.03]  &  [58.2 $\pm$ 0.6]    &  1.481  $\pm$ 0.023  &  0.919  $\pm$ 0.017  &  \citet{Fukagawa:2009iq}  \\
2004 Jul 14  &  2004.534  &  Keck/NIRC2 &  [1.716  $\pm$  0.007]  &  [59.0 $\pm$ 0.17]    &  1.471  $\pm$ 0.005  &  0.884  $\pm$ 0.005  &  \citet{Marois:2008ei}  \\
2005 Jul 15  &  2005.534  &  Keck/NIRC2 &  [1.724  $\pm$  0.007]  &  [60.2 $\pm$ 0.17]    &  1.496  $\pm$ 0.005  &  0.856  $\pm$ 0.005  &  \citet{Currie:2012ea}  \\
2007 Aug 02  &  2007.583  &  Keck/NIRC2 &  [1.726  $\pm$  0.004]  &  [61.8 $\pm$ 0.1]    &  1.522  $\pm$ 0.003  &  0.815  $\pm$ 0.003  &  \citet{Metchev:2009jl}  \\
2007 Aug 02\tablenotemark{a}  & 2007.583  &  Keck/NIRC2 &  [1.721 $\pm$ 0.004]     &  [60.90 $\pm$ 0.10]    &  1.504 $\pm$ 0.003   &  0.837 $\pm$ 0.003 & \citet{Konopacky:2016vk} \\
2007 Oct 25  &  2007.813  &  Gemini-N/NIRI  &  [1.713  $\pm$  0.007]  &  [62.0 $\pm$ 0.17]    &  1.512  $\pm$ 0.005  &  0.805  $\pm$ 0.005  &  \citet{Marois:2008ei}  \\
\enddata
\tablecomments{Astrometry listed in brackets either in $\rho$ and $\theta$ or $\Delta$R.A. and $\Delta$Dec is inferred from published measurements.}
\tablenotetext{a}{Re-reduction of published data.}
\tablenotetext{b}{\citet{Konopacky:2016vk} note that this epoch of astrometry for HR~8799~d may be biased because the primary is positioned near the edge of the coronagraph.}
\tablenotetext{c}{Sum of observations taken on 2009 Nov 24, Nov 25, and Dec 17.}
\tablenotetext{d}{\citet{Bonnefoy:2011ex} report astrometry using two reduction procedures for each data set.  Values listed here are the weighted average of both methods.}
\end{deluxetable}
\clearpage

\clearpage
\newpage

\end{document}